\documentclass[12pt]{article}
\usepackage{epsfig}
\begin{document}
\newcommand{\be}{\begin{equation}}
\newcommand{\ee}{\end{equation}}
\newcommand{\bea}{\begin{eqnarray}}
\newcommand{\eea}{\end{eqnarray}}
\newcommand{\rotatefigure}[2]
\centerline{\bf  Exact resolution of the Baxter
equation for reggeized gluon interactions}
\begin{center}
H. J. de Vega\\
LPTHE\footnote{Laboratoire Associ\'{e} au CNRS, UMR 7589.},
Universit\'e Pierre et Marie Curie (Paris VI)\\ 
et Denis Diderot (Paris 7), Tour 16, 1er. \'etage,\\
4, Place Jussieu, 75252 Paris, cedex 05, France \\
\medskip
L. N. Lipatov \footnote{Supported by the RFFI and INTAS grants:
1997-31696, 2000-366}\\ 
Petersburg Nuclear Physics Institute,\\
Gatchina, 188300, St. Petersburg, Russia and \\
LPMT Universit\'e Montpellier 2, \\
Place Eug\`ene Bataillon, 34095 Montpellier Cedex 05, France
\end{center}
\vskip 15.0pt
\date{\today}
\centerline{\bf Abstract}
\noindent
The interaction of reggeized gluons in multi-colour QCD is
considered in the
Baxter-Sklyanin representation, where the wave function is expressed as a
product of Baxter functions $Q(\lambda)$ and a pseudo-vacuum state. We
find $n$ solutions of the Baxter equation for a composite state of $n$
gluons with  poles of  rank $r$ in the upper $\lambda$ semi-plane and of
rank $n-1-r$ in the lower $\lambda$ semi-plane ($0 \leq r \leq n-1$).
These solutions are related by $n-2$ linear equations with
coefficients depending on $\coth (\pi \lambda)$. The poles cancel in
the wave function, bilinear combination of holomorphic and
anti-holomorphic Baxter functions, guaranteeing its normalizability.
The quantization of the intercepts of the corresponding Regge singularities
appears as a result of the physical requirements that the holomorphic
energies for all solutions of  the Baxter equation are the same and the
total energies,
calculated around two singularities $\lambda , \lambda ^*
\rightarrow \pm i$, coincide. It
results in simple properties of the zeroes of the Baxter
functions. For illustration we calculate the parameters of the
reggeon states constructed from three and four gluons. For the 
Odderon the ground state has conformal spin
$ |m -\widetilde{m}|=1 $ and its intercept equals unity. The ground
state of four reggeized gluons possesses conformal spin $2$ and its
intercept turns out to be higher than that for the BFKL Pomeron.
We calculate the anomalous dimensions of
the corresponding operators for arbitrary $\alpha _s/\omega$.  

\section{Introduction}

The leading logarithmic asymptotics (LLA) of  scattering amplitudes in the
Regge limit of high energies $\sqrt{s}$
and fixed momentum transfers $q=\sqrt{-t}$  is obtained by
calculating and summing all contributions $\left( g^{2}\ln s\right)
^{r}$, where $g$ is the QCD coupling constant. In this approximation
the BFKL Pomeron is a composite state of two
reggeized gluons \cite {BFKL}. The BFKL equation for the
Pomeron wave function is closely related to the DGLAP equation
for the parton distributions \cite{DGLAP}.
Next-to-leading corrections to its
kernel were calculated in QCD \cite{FL}  and in
supersymmetric gauge theories \cite{KoLi}.

The asymptotic behaviour $\propto s^{j_{0}}$ of scattering amplitudes is
governed by the $j$-plane singularities of the $t$-channel partial waves
$f_{j}(t)$
\begin{equation}
A(s,t)=\int_{\sigma -i\infty }^{\sigma +i\infty }\frac{dj}{2\pi i} \; \xi
_{j}\,\,s^{j}\,f_{j}(t)\sim \xi _{1+\omega _{0}\,}s^{1+\omega _{0}}\,.
\end{equation}
Here the contour of integration in $j$ is situated to the right of the
leading singularity $j_{0}=1+\omega _{0}$ of $f_{j}(t)$ ($j_{0}<\sigma$)
and the signature factor $\xi _{j}$ is $\simeq i^{n-1}$ for the
$t$-channel exchange of $n$-reggeized gluons. Its intercept $\omega _{0}$ 
is proportional to the
ground state energy $E_{0}$ of the Schr\"{o}dinger-like equation
\cite{BFKL,BKP}:
 \begin{equation}
Hf=Ef\,\,,\,\omega _{0}=-\frac{g^{2}}{8\pi ^{2}}\,N_{c}\,E_{0}\,.
\end{equation}
The wave function $f$ depends on the two-dimensional impact
parameters $\overrightarrow{\rho _k}$ - positions of the reggeized
gluons. It is convenient to
introduce the holomorphic ($\rho _k=x_k+iy_k$) and anti-holomorphic
($\rho _k^*=x_k-iy_k$) coordinates and their corresponding momenta
$p_k=i\frac{\partial}{\partial \rho _k}$ and $p^*_k=
i\frac{\partial}{\partial \rho  ^*_k}$.

In multicolour QCD $N_{c}\rightarrow \infty $ the colour
structure of the BFKL equation in LLA is significantly simplified.
As a result, each reggeized gluon
interacts only with its two neighbours \cite{separ}:
\begin{equation}
H\,=\,\frac{1}{2}\sum_{k=1}^{n}\,H_{k,k+1}\,.
\end{equation}
Note, that for three gluon composite state, describing the Odderon
responsible for the high energy behaviour of
the differences of total cross-sections for particle-particle and
particle-anti-particle interactions, this
simplification is valid for arbitrary $N_c$ \cite{BKP,odd}.

The Hamiltonian $H$  has the properties of the M\"{o}bius invariance
\cite{conf} and of the holomorphic separability
\cite{separ}:

\begin{equation}
H=\frac{1}{2}(h+h^{*}),\,\,\,\left[ h,h^{*}\right] =0\,,
\end{equation}
where the holomorphic and anti-holomorphic Hamiltonians

\begin{equation}
h=\sum_{k=1}^{n}h_{k,k+1\,},\,h^{\ast }=\sum_{k=1}^{n}h_{k,k+1\,}^{\ast }
\end{equation}
are expressed in terms of the BFKL operator \cite{intmot}
\begin{equation}
h_{k,k+1}=\log (p_{k})+\log (p_{k+1})+\frac{1}{p_{k}}\log (\rho
_{k,k+1}) \; p_{k}+\frac{1}{p_{k+1}}\log(\rho _{k,k+1}) \;
p_{k+1}+2\,\gamma \,. 
\end{equation}
Here $\rho _{k,k+1}=\rho _{k}-\rho _{k+1}$ and
$\gamma =-\psi (1)$ is the Euler-Mascheroni  constant.

The wave function $f_{m,\widetilde{m}}(\overrightarrow{\rho _{1}},
\overrightarrow{\rho _{2}},...,\overrightarrow{\rho
_{n}};\overrightarrow{
\rho _{0}})$ of the colourless composite state described by the operator
$O_{m\widetilde{,m}}(\overrightarrow{\rho _{0}})$  belongs to the
principal series of unitary representations of the M\"{o}bius group \cite
{conf}. For these representations the conformal weights
\begin{equation}
m=1/2+i\nu +n/2 \quad , \quad \widetilde{m}=1/2+i\nu -n/2
\end{equation}
are expressed in terms of the anomalous dimension $\gamma =1/2+i\nu $
of $O_{m\widetilde{,m}}(\overrightarrow{\rho _{0}})$  and
its integer conformal spin $n$. Furthermore, the
eigenvalues of two Casimir operators $M^2$ and $M^{*2}$ of the M\"{o}bius
group are equal to $m(m-1)$ and $\widetilde{m}(\widetilde{m}-1)$,
respectively.

Owing to the holomorphic separability of $H$, the wave function
has the property of  holomorphic factorization \cite{separ}:
\begin{equation}
f_{m,\widetilde{m}}(\overrightarrow{\rho _{1}},\overrightarrow{\rho _{2}}
,...,\overrightarrow{\rho _{n}};\overrightarrow{\rho _{0}}
)=\sum_{r,l}c_{r,l}\,f_{m}^{r}(\rho _{1},\rho _{2},...,\rho _{n};\rho
_{0})\,f_{\widetilde{m}}^{l}(\rho _{1}^{*},\rho _{2}^{*},...,\rho
_{n}^{*};\rho _{0}^{*})\,,
\end{equation}
where $r$ and $l$ enumerate the different solutions of the Schr\"{o}dinger
equations in the holomorphic and anti-holomorphic sub-spaces:
\begin{equation}
\epsilon _{m}\,f_{m}=h\,f_{m} \quad , \quad \epsilon _{\widetilde{m}}\,f_{
\widetilde{m}}=h^{\ast }\,f_{\widetilde{m}} \quad , \quad
E_{m,\widetilde{m}}=\epsilon _{m}+\epsilon _{\widetilde{m}}\,.
\end{equation}
Similarly to the case of two-dimensional conformal field theories, the
coefficients $c_{r,l}$ are obtained imposing
single-valuedness to
$f_{m,\widetilde{m}}(\overrightarrow{\rho _{1}},
\overrightarrow{\rho _{2}},...,\overrightarrow{\rho
_{n}};\overrightarrow{
\rho _{0}})$ as a function of
the two-dimensional variables $\overrightarrow{\rho}_i$.

There are two different normalization conditions for the wave function
\cite{intmot}:
\begin{equation}
\left\| f\right\| _{1}^{2}=\int \prod_{r=1}^{n}d^{2}\rho _{r}\,\left|
\prod_{r=1}^{n}\rho _{r,r+1}^{-1}\,\,f\right| ^{2},\,\,\,\left\|
f\right\|
_{2}^{2}=\int \prod_{r=1}^{n}d^{2}\rho _{r}\left|
\prod_{r=1}^{n}p_{r}\,\,f\right| ^{2}
\end{equation}
compatible with the hermiticity  of $H$.
This property is related with the fact \cite{intmot}, that $h$ commutes
with the differential operator
\begin{equation}
A=\rho _{12}\rho _{23}...\rho _{n1}\,p_1p_2...p_n\,.
\end{equation}
Furthermore \cite{integr}, there is a family $\{q_{r}\}$\thinspace of
mutually commuting differential operators being the integrals of motion:
\begin{equation}
\left[ q_{r},q_{s}\right] =0 \quad ,\quad \left[ q_{r},h\right] =0.
\end{equation}
They can be obtained by the small-$u$ expansion  of the transfer matrix
for
the $XXX$
model \cite{integr}
\begin{equation}
T(u)=tr\,\left[
L_{1}(u)L_{2}(u)...L_{n}(u)\right]=\sum_{r=0}^{n}u^{n-r}\,q_{r}, 
\end{equation}
where the $L$-operators are
\begin{equation}
L_{k}(u)=\left(
\begin{array}{cc}
u+\rho _{k0}\,p_{k} & -p_{k} \\
\rho _{k0}^{2}\,p_{k} & u-\rho _{k0}\,p_{k}
\end{array}
\right) =\left(
\begin{array}{cc}
u & 0 \\
0 & u
\end{array}
\right) +\left(
\begin{array}{c}
1 \\
\rho _{k0}
\end{array}
\right) \left(
\begin{array}{cc}
- \rho _{k0} & 1
\end{array}
\right) p_{k}\,.
\end{equation}
In particular $q_{n}$ is equal to $A$ and $q_{2}$ is proportional to
$M^{2}$.

The transfer matrix is the trace of the monodromy matrix $t(u)$:
\begin{equation}
T(u)=tr\,[t(u)] \quad , \quad t(u)=L_{1}(u)L_{2}(u)...L_{n}(u)\,.
\end{equation}
It can be shown \cite{integr,fadd}, that $t(u)$ satisfies the
Yang-Baxter equation:
\begin{equation}
t_{r_{1}^{\prime }}^{s_{1}}(u)\,t_{r_{2}^{\prime
}}^{s_{2}}(v)\,l_{r_{1}r_{2}}^{r_{1}^{\prime }r_{2}^{\prime
}}(v-u)=l_{s_{1}^{\prime }s_{2}^{\prime
}}^{s_{1}s_{2}}(v-u)\,t_{r_{2}}^{s_{2}^{\prime
}}(v)\,t_{r_{1}}^{s_{1}^{\prime }}(u)\,,
\end{equation}
where $l(w)$ is the $L$-operator for the well-known Heisenberg spin chain:
\begin{equation}
l_{s_{1}^{\prime }s_{2}^{\prime }}^{s_{1}s_{2}}(w)=w\,\delta
_{s_{1}^{\prime
}}^{s_{1}}\,\delta _{s_{2}^{\prime }}^{s_{2}}+i\,\delta _{s_{2}^{\prime
}}^{s_{1}}\,\delta _{s_{1}^{\prime }}^{s_{2}}\,.
\end{equation}

\section{Baxter-Sklyanin representation}

Thus, the problem of finding  solutions of the Schr\"odinger equation
for the reggeized gluon interaction reduces to the search of a
representation for the monodromy matrix satisfying  the Yang-Baxter
bilinear relations \cite{integr}. For this purpose  the algebraic
Bethe Ansatz is appropriate \cite{fadd}. It is important, that the reggeon
Hamiltonian in the multi-colour QCD coincides with the local Hamiltonian
of the integrable Heisenberg model with the spins being the generators
of the non-compact M\"{o}bius group  $SL(2,C)$ \cite{Heis, fadkor}. For 
the case of three
gluons the wave function and intercepts of the corresponding Regge 
singularities were found in ref.\cite{JW} with the use the integral of 
motion discovered in ref.\cite{intmot}.

The integrals of motion and the hamiltonian for $n$ reggeized gluons have 
an additional symmetry under the transformation \cite{dual}
\begin{equation}
p_{k}\rightarrow \rho _{k,k+1}\rightarrow p_{k+1}\,,
\end{equation}
combined with the operator transposition. This duality symmetry allows
to relate the wave function of a composite state with the Fourier
transformed wave function of 
(generally) another physical state. In particular, it
gives a possibility to construct a new Odderon solution having the
intercept exactly equal to unity \cite{new}. The duality symmetry can
be interpreted as a symmetry among the states constructed from the
reggeons with positive and negative signatures
\cite{new}.  Indeed, the Regge trajectories for these states with
gluon quantum numbers are degenerated in multi-colour QCD.
 
In the framework of the Bethe Ansatz it is convenient to work in
the conjugated space \cite{fadkor}, where the monodromy matrix is
parametrized as follows,
\begin{equation}
\widetilde{t}(u)=\widetilde{L}_{n}(u)...\widetilde{L}_{1}(u)=\left(
\begin{array}{cc}
A(u)\, & B(u) \\
C(u) & D(u)
\end{array}
\right) \,.
\end{equation}
Here $\widetilde{L}_{k}(u)$ is given by
\begin{equation}
\widetilde{L}_{k}(u)=\left(
\begin{array}{cc}
u+p_{k}\rho _{k0}\, & -p_{k}\, \\
p_{k}\rho _{k0}^{2}\, & u-p_{k}\rho _{k0}\,
\end{array}
\right) \,.
\end{equation}
Now the equation for the pseudo-vacuum state
\begin{equation}
C(u)\,|0\rangle ^{t}=0
\end{equation}
has the following solution \cite{fadkor}
\begin{equation}
|0\rangle ^{t}=\prod_{k=1}^{n}\frac{1}{\rho _{k0}^{2}}\,.
\end{equation}
In the total impact parameter space $\overrightarrow{\rho}$ the
pseudo-vacuum wave function,
\begin{equation}
\Psi ^{(0)}(\overrightarrow{\rho _{1}},\,\overrightarrow{\rho
_{2}},\,\ldots
,\overrightarrow{\rho _{n}}; \overrightarrow{\rho _{0}})
=\prod_{k=1}^{n}\frac{1}{|\rho
_{k0}|^{4}}\,.\,\,
\end{equation}
is an eigenfunction of the transfer matrix,
\begin{equation}
\left[ A(u)+D(u)\right] |0\rangle ^{t}=\left[
(u-i\,)^{n}+(u+i\,)^{n}\right]\,|0\rangle ^{t}.
\end{equation}
The pseudo-vacuum state does not belong to the principal series of
unitary representations because it has the conformal weight $m=n$.

A powerful approach to construct the physical states in the framework of
the Bethe Ansatz is based on the use of
the  Baxter equation for the Baxter function $Q(\lambda)$ \cite{bax, sklya}.
The Baxter equation for the $n$-reggeon composite states can be written as
follows (see \cite{fadkor, hect, dkm})
\begin{equation}
\Lambda ^{(n)}(\lambda ;\;\vec{\mu})\,
Q\left( \lambda ;\,m,\vec{\mu}\right)=
(\lambda +i)^n \; Q\left(
\lambda +i ;\,m, \vec{\mu}\right) +(\lambda -i)^n \; Q\left(
\lambda
- i ;\,m,\vec{\mu}\right)
\; ,
\end{equation}
where $\Lambda ^{(n)}(\lambda )$ is the polynomial
\begin{equation}
\Lambda ^{(n)}(\lambda ;\;\vec{\mu})=\sum _{k=0}^n  (-i)^k\,\mu
_k\,\lambda
^{n-k}\,,\;\mu _0=2,\; \mu _1=0,\;\mu _2=m(m-1)\,.
\end{equation}
Here we assume in accordance with ref.\cite{hect}, that the quantities
$\mu  _k=i^k\,q_k$ are real which is compatible with the
single-valuedness condition of the wave functions [$q_k$
stand here for the eigenvalues of the integrals of motion].

The eigenfunctions of the holomorphic Schr\"{o}dinger equation can be
expressed through the Baxter function $Q(\lambda )$ using the
Sklyanin Ansatz \cite{sklya}:
\begin{equation}
f(\rho _{1},\rho _{2},...,\rho _{n}; \rho _{0})=Q(\widehat{\lambda
}_{1} ;\,m,\vec{\mu})\,Q(\widehat{\lambda
}_{2} ;\,m,\vec{\mu})...Q(\widehat{\lambda }_{n-1}
;\,m,\vec{\mu})|0\rangle ^{t}\,.
\end{equation}
where $\widehat{\lambda }_{r}$ are the operator zeroes of the matrix
element
$B(u)$ of the monodromy matrix:
\begin{equation}
B(u)=-P\,\prod_{r=1}^{n-1}(u-\widehat{\lambda }_{r})\,,\,\,P=
\sum_{k=1}^{n}p_{k}\,.
\end{equation}
These expressions are well defined because the operators
$\widehat{\lambda }_{r}$ and $P$  commute   \cite{sklya}
\begin{equation}
\left[ \widehat{\lambda }_{r}\,,\widehat{\lambda }_{s}\right] =\left[
\widehat{\lambda }_{r}\,,P\right] =0.
\end{equation}
In ref.\cite{fadkor} it was  assumed without any convincing arguments,
that  $Q(\lambda )$ is
an entire function in the complex $\lambda$ plane. One of the purposes
of our previous paper \cite{hect} was to find the class of functions to
which the Baxter functions belong.  For this purpose we performed an unitary
transformation of the wave
function of the composite state of $n$ reggeized gluons from the
coordinate representation to the Baxter-Sklyanin representation
in which the operators $\widehat{\lambda _r}$ are diagonal
\cite{hect} (see also \cite{dkm}). The kernel of this
transformation was expressed through the eigenfunctions of
the operators $B(u)$ and $B^*(u)$. For the cases of the Pomeron
and Odderon the unitary transformation was constructed in an explicit
form \cite{hect}.
As a consequence of the single-valuedness condition for the kernel of
this transformation one
obtains the quantization of the
arguments of the Baxter functions $Q(\lambda)$ and $Q(\lambda ^*)$
in the holomorphic and anti-holomorphic sub-spaces (see \cite{hect, dkm}):
\begin{equation}
\lambda =\sigma +i \; \frac{N}{2} \quad , \quad
\lambda ^* =\sigma -i\; \frac{N}{2} \quad ,
\end{equation}
where $\sigma$ and $N$ are real and integer numbers, respectively.

In ref.\cite{hect} we proposed a general method of solving the Baxter
equation for the $n$-reggeon composite state. To begin with,
the simplest $n$-reggeon solution of this equation is searched in the
form of a sum over the poles of  orders from $1$ up to $n-1$ situated in
the upper semi-plane
\begin{equation}
Q^{(n-1)}\left( \lambda ;\,m,\vec{\mu}\right)=\sum _{r=0}^\infty
\frac{P^{(n-2)}_{r;m,\vec{\mu}}(\lambda )}{(\lambda
- i\,r)^{n-1}} \; ,
\end{equation}
where the $ P^{(n-2)}_{r;m,\vec{\mu}}(\lambda ) $ are polynomials in $
\lambda $ of degree $ n-2 $.
Inserting  this Ansatz in the Baxter equation leads to  recurrence relations
for the polynomials $P^{(n-2)}_{r;m,\vec{\mu}}(\lambda )$, which
allows us to calculate them successively starting from
$P^{(n-2)}_{0;m,\vec{\mu}}(\lambda ) $ \cite{hect}.

One can normalize this solution imposing the constraint
\begin{equation}
\lim _{\lambda \rightarrow 0}P^{(n-2)}_{0;m,\vec{\mu
}}(\lambda )=1
\end{equation}
Then, the remaining coefficients of the polynomial
$P^{(n-2)}_{0,m,\vec{\mu }}(\lambda )$ are calculated from the
condition
\begin{equation}
\lim _{\lambda \rightarrow \infty} Q ^{(n-1)}\left( \lambda
;\,m,\vec{\mu}\right) \sim \lambda ^{-n+m}\,
\end{equation}
which is a necessary condition for $ Q^{(n-1)}\left( \lambda
;\,m,\vec{\mu}\right)$ to be a solution of the Baxter equation at
$\lambda \rightarrow \infty$. It is enough to require
\begin{equation}
\lim_{\lambda \rightarrow \infty} \lambda ^{n-2} \,\sum
_{r=0}^\infty
{P^{(n-2)}_{r;m,\vec{\mu}}(\lambda ) \over (\lambda -i\,r)^{n-1}}
=0 \,.
\end{equation}
This condition gives $n-1$ linear equations allowing to
calculate all coefficients of the polynomial $P^{(n-2)}_{0,m,\vec{\mu
}}(\lambda)$.

The existence of the second independent solution
\begin{equation}
Q^{(0)}\left( \lambda ;\,m,\vec{\mu}\right)=
Q^{(n-1)}\left( -\lambda ;\,m,\vec{\mu^s}\right)=\sum _{r=0}^\infty
\frac{P^{(n-2)}_{r;m,\vec{\mu ^s}}(-\lambda )}{(-\lambda
- i\,r)^{n-1}}\,,
\end{equation}
where 
$$
\mu ^s _k=(-1)^k \mu _k\,,
$$
is related with the invariance of the Baxter equation under the
simultaneous transformations
$$
\lambda \rightarrow -\lambda \,,\,\,\mu \rightarrow \mu ^s \,.
$$
One can verify \cite{hect} that
$$
 Q^*\left( -\lambda ;\,m,\vec{\mu^s}\right)=Q\left(
\lambda ^*;\,\widetilde{m},\vec{\mu^s}\right)\,.
$$

It turns out \cite{hect}, that there is a set of the Baxter functions
$Q^{(t)}$ ($t=0, 1,...,n-1$) having  poles simultaneously in the upper
and lower half-$\lambda$ planes.
$$
Q^{(t)}\left( \lambda ;\,m,\vec{\mu}\right)=\sum _{r=0}^\infty
\left[ \frac{P^{(t-1)}_{r;m,\vec{\mu}}(\lambda )}{(\lambda
- i\,r)^t}+\frac{P^{(n-2-t)}_{r;m,\vec{\mu ^s}}(-\lambda )}{(-\lambda
- i\,r)^{n-1-t}}\right]\,,
$$
where the polynomials $P^{(t-1)}_r$ and $P^{(n-2-t)}_r$
are fixed by the recurrence relations following
from the Baxter equation and from the condition that the above Baxter
functions decrease at infinity  more rapidly than
$\lambda ^{-n+2}$. These solutions $Q^{(t)}$ are linear combinations of
$Q^{(n-1)}\left( \lambda ;\,m,\vec{\mu }\right)$ and $Q^{(n-1)}\left(
- \lambda
;\,m,\vec{\mu^s}\right)$
with the coefficients depending on $\coth (\pi \lambda)$ \cite{hect}.

Using all these functions in the
holomorphic and anti-holomorphic spaces one can  construct 
the normalizable Baxter function
$Q_{m,\,\widetilde{m},\,\vec{\mu}} \left( \overrightarrow{\lambda}\right)$ 
in the space ($\sigma , N$) without poles at $\sigma =0$ \cite{hect}
\begin{equation}
Q_{m,\,\widetilde{m},\,\vec{\mu}}
\left( \overrightarrow{\lambda}\right)=\sum _{t,l}C_{t,l}\,
Q^{(t)}\left( \lambda ;\,m,\vec{\mu}\right) \,Q^{(l)}\left(
\lambda ^*;\,\widetilde{m},\vec{\mu  ^s}\right)\,
\end{equation}
by adjusting  for this purpose the coefficients $C_{t,l}$.

Another problem in the Baxter approach is the calculation of the
energy $E_{m,\widetilde{m}}$, because the expression suggested in ref.
\cite{fadkor} leads to an infinite result for meromorphic Baxter
functions. This problem was solved in ref.\cite{hect} by an unitary
transformation of the Hamiltonian to the Baxter-Sklyanin representation.

In the region, where the
gluon momenta  are strongly ordered in their values,
$$
|p_n|<<|p_{n-1}|<<...<<|p_1|=1
$$
the unitary transformation of the wave function $\Psi _{m,\widetilde{m}}$
to the Baxter-Sklyanin representation
is
significantly simplified \cite{hect}

\begin{equation}
\Psi_{m,\widetilde{m}}(\overrightarrow{p_1},
...\,,
\overrightarrow{p_n}) \sim \prod _{k=1}^{n-1}
\left( \int_{-\infty}^{+\infty} d \sigma _k
\sum _{N_k=-\infty}^{+\infty}  \,
 p_{k-1}^{i\lambda _{k}^{*}} \; p_{k-1}^{*i\lambda_{k}}
\right)\Psi
_{m,\widetilde{m}}(
\overrightarrow{\lambda _1},...,\overrightarrow{\lambda}_{n-1})\,.
\end{equation}
On the other hand, in this region
\begin{equation}
\Psi
_{m,\widetilde{m}}(\overrightarrow{p_1},\ldots \,
, \overrightarrow{p_n})\sim c_n \, \left| p_n\right| ^2\ln
\frac{1}{\left| p_n\right| ^2} \,,
\end{equation}
where $c_n$ is a constant \cite{hect}. Therefore $\Psi_{m,\widetilde{m}}(
\overrightarrow{\lambda _1},\ldots,\overrightarrow{\lambda}_{n-1})$
has the first order poles at $\lambda _{n-1}=i$ and $\lambda
_{n-1}^*=i$ for $N_{n-1}=0$ and the pole singularities at $\lambda
_k=\lambda
_k^*=0$ ($k=1,2,...,n-2$). The action of the Hamiltonian $H$
on the function $\Psi_{m,\widetilde{m}}$ near these
singularities after the unitary transformation to the $\lambda$-space is
drastically simplified \cite{hect}. Moreover, using the
Sklyanin factorized expression for the wave function
$\Psi_{m,\widetilde{m}}(
\overrightarrow{\lambda _1},\ldots,\overrightarrow{\lambda}_{n-1})$
we express the energy in terms of the behaviour of
$Q(\overrightarrow{\lambda _{n-1}})$
near the pole at $\lambda _{n-1}=i, \,\lambda ^*_{n-1}=i$
and obtain \cite{hect}
\begin{equation}
E=i\lim_{\lambda ,\lambda ^{\ast }\rightarrow i}\frac{\partial }{\partial
\lambda }\frac{\partial }{\partial \lambda ^{\ast }}\ln \left[ (\lambda
- i)^{n-1}(\lambda ^{\ast }-i)^{n-1}\left| \lambda \right| ^{2\,n}\,
 Q_{m,\,\widetilde{m},\,\vec{\mu}}
\left( \overrightarrow{\lambda}\right)\right] \,.
\end{equation}
It is important, that the consideration of the kinematical region with
an opposite ordering of the gluon momenta $p_k$ leads to an
analogous expression for $E$ in terms of
the behaviour of $ Q_{m,\,\widetilde{m},\,\vec{\mu}}
\left( \overrightarrow{\lambda}\right)$ near $\lambda , \lambda ^*= -i$.
The coincidence of these energies imposes a constraint on the spectrum
of the integrals of motion $q_k$ ($k=3,4,...,n$).
 
Since the function $Q_{m,\,\widetilde{m},\,\vec{\mu}}
\left( \overrightarrow{\lambda}\right)$ is a bilinear
combination of the independent Baxter functions $Q^{(t)}(\lambda )$ and
$Q^{(l)}(\lambda ^*)$ ($t\,,l\,=1,2,...,n$),
the holomorphic (anti-holomorphic) energies
for all solutions should be the same
\begin{equation}
\epsilon _m =i\lim_{\lambda \rightarrow i}\frac{\partial
}{\partial \lambda }\ln \left[\lambda
^n\,P^{(t-1)}_{1;m,\vec{\mu}}(\lambda )
\right]\,.
\end{equation}
This leads to the quantization of the integrals of motion $q_k$
\cite{hect}.

Note, that this condition  can be obtained also as one of the consequences
of the absence of poles in $ Q_{m,\,\widetilde{m},\,\vec{\mu}}
\left( \overrightarrow{\lambda}\right)$ at $\sigma
=0$ for $|N|>0$. Because the products of poles $(\lambda -ir)^{-s}$ and
$(-\lambda^* - ir')^{s'}$ appear in the corresponding bilinear combinations,
to cancel all singularities one should
expand the functions $Q^{(t)},\,Q^{(l)}$ in Laurent series near these
poles and keep all relevant terms. The
coefficients of the Laurent series satisfy recurrence relations
similar to those for the residues of the poles. However, it turns out that
they {\bf are not} proportional to the residues of the poles.. 
Even a special choice of the values for the integrals of motion $q_r$
does not lead to such proportionality because these recurrence relations
are different at $k=2$ \cite{hect}.

We discuss below the solution of the Baxter equation by the method
proposed in ref.\cite{hect}. Note, that for $n=3$ this method gives
results in full agreement with those obtained by other approaches
\cite{JW,dual}. We find in the present paper the wave functions and
intercepts for the quarteton (four reggeons state). 

\section{Meromorphic solutions of the Baxter equation}
Let us rewrite the Baxter equation in a real form introducing the new
variable $x \equiv - i \lambda$,
\begin{equation}\label{Breal}
\Omega (x, \vec{\mu})\,Q(x,\vec{\mu})=(x+1)^n\,Q(x+1,\vec{\mu})+
(x-1)^n\,Q(x-1,\vec{\mu}) \,,
\end{equation}
where
\begin{equation}
\Omega (x, \vec{\mu})=\sum _{k=0}^{n}\, (-1)^k\,\mu _k \,x^{n-k} \,
\end{equation}
and
$$
\mu _0 =2\,,\,\,\mu _1 = 0 \,,\,\, \mu _2 =m(m-1) \,,
$$
assuming that the eigenvalues of the integrals of motion $\mu _k$ ($k>2$)
are real numbers.

Then, the solution of the Baxter equation possessing poles only at
$ x = l=0,1,2,...$ can be written as follows,
\begin{equation}
Q^{(n-1)}(x,\vec{\mu})=\sum _{l=0}^{\infty} \,\left[
\frac{a_{l}(\vec{\mu})}{(x-l)^{n-1}}+\frac{b_{l}(\vec{\mu})}{(x-l)^{n-2}}+
...+\frac{z_{l}(\vec{\mu})}{x-l}\,\right] \,,
\end{equation}
where the residues
$a_{l}(\vec{\mu}),\,b_{l}(\vec{\mu}),...,\,z_{l}(\vec{\mu})$ satisfy
recurrence relations which can be easily obtained from the Baxter
equation and its derivatives in the limit $x \rightarrow l$.
Using these relations we can express all these residues in terms
of $a_{0}(\vec{\mu}),\,b_{0}(\vec{\mu}),...,\,z_{0}(\vec{\mu})$.
In addition, imposing the Baxter equation at $x \rightarrow \infty$,
\begin{equation}\label{condasi}
\lim _{x \rightarrow \infty} \; x^s \; Q^{(n-1)}(x,\vec{\mu}) =0
\quad ,\quad s=1,2, \ldots ,n-2\,,
\end{equation}
fixes the parameters $ b_{0}(\vec{\mu}),...,\,z_{0}(\vec{\mu})$ in
terms of $a_{0}(\vec{\mu})$. We then require the normalization  condition
$$
a_{0}(\vec{\mu})=1 \; ,
$$
without losing generality. In this normalization the holomorphic
energy is (see \cite{hect})
$$
\epsilon _n =\frac{b_1}{a_1}+n=b_{0}-\frac{\mu _{n-1}}{\mu _n}\,.
$$
Thus, the solution $Q^{(n-1)}(x,\vec{\mu})$ is uniquely defined and
can be explicitly constructed.

Let us further introduce a set of the auxiliary functions for
$r=1,2,...,n-1$
\begin{equation}\label{efe}
f_r(x,\vec{\mu})=\sum _{l=0}^{\infty} \,\left[
\frac{\widetilde{a}_{l}(\vec{\mu})}{(x-l)^{r}}+
\frac{\widetilde{b}_{l}(\vec{\mu})}{(x-l)^{r-1}}+
...+\frac{\widetilde{g}_{l}(\vec{\mu})}{x-l}\,\right] \,,
\end{equation}
where the coefficients $\widetilde{a}_{l},...,\widetilde{g}_l$ satisfy
the same recurrent relations as $a_{l},...,z_l$ but with other initial
conditions
\begin{equation}
\widetilde{a}_{0}=1,\,\widetilde{b}_{0}=...=\widetilde{g}_0=0 \,.
\end{equation}
Note, that all functions $f_r(x,\vec{\mu})$ are expressed in terms of a
subset of pole residues $\widetilde{a}_{l},...,\widetilde{z}_l$ for
$f_{n-1}(x,\vec{\mu})$.

Now we write the Baxter function $Q^{(n-1)}(x,\vec{\mu})$ as a linear
combination of $f_r(x,\vec{\mu})$
\begin{equation}\label{solnm1}
Q^{(n-1)}(x,\vec{\mu})=\sum
_{r=1}^{n-1}C_r(\vec{\mu})\,f_r(x,\vec{\mu})\,.
\end{equation}

In order to impose the asymptotic condition (\ref{condasi}) on $
Q^{(n-1)}(x,\vec{\mu}) $, we use the binomial series for the terms in
eq.(\ref{efe}),
$$
\frac{1}{(x-l)^{j}} = \sum_{n=0}^{\infty} \frac{l^n}{x^{n+j}}
\frac{(j+n-1)!}{n! \; (j-1)!} =\sum _{k=j}^{\infty}
\frac{l^{k-j}}{x^k}\frac{(k-1)!}{(k-j)!(j-1)!}\; .
$$
We  find a set of $n-2$ linear equations on the  coefficients
$C_r(\vec{\mu})$:
$$
\sum _{r=1}^{n-1}G_{k,r}(\vec{\mu}) \; C_r(\vec{\mu})=0 \quad ,\quad
k=1,2, \ldots ,n-2 \quad \quad C_{n-1}(\vec{\mu})=1\,\,,
$$
where
\begin{equation}\label{defG}
G_{k,r}(\vec{\mu})=\sum _{l=0}^{\infty} \,\left[
\widetilde{a}_{l}(\vec{\mu})\; \frac{(k-1)! \; l^{k-r}}{
(k-r)!\; (r-1)!}+\widetilde{b}_{l}(\vec{\mu})\; \frac{
(k-1)! \; l^{k-r+1}}{(k-r+1)!\; (r-2)!}+ \ldots
+\widetilde{g}_{l}(\vec{\mu})\; l^{k-1}\right] \; .
\end{equation}
Obviously, the symmetric solution
\begin{equation}
Q^{(0)}(x,\vec{\mu})=Q^{(n-1)}(-x,\vec{\mu^s}) \quad , \quad \mbox{where} \quad
\mu_r^s \equiv (-1)^r \mu _r  \; , 
\end{equation}
can be constructed in a similar way.  It has poles at $x=-l$ ($l=0,1,...$).
 
But there are other `minimal' solutions
$Q^{(t)}(x,\vec{\mu})$ ($t=1,2,...,n-2 $) of the Baxter equation having
$t$-order poles at positive integer $x$ and $(n-1-t)$-order poles at
negative
integer $x$ \cite{hect}
\begin{equation}\label{solt}
Q^{(t)}(x,\vec{\mu})=\sum
_{r=1}^{t}C_r^{(t)}(\vec{\mu})\,f_r(x,\vec{\mu})+
\beta ^{(t)}(\vec{\mu})\,\sum
_{r=1}^{n-1-t}C_r^{(n-1-t)}(\vec{\mu^s})\,f_r(-x,\vec{\mu^s})\,,
\end{equation}
where the meromorphic functions $f_r(x,\vec{\mu})$ were defined above.
Such form of the solution is related by the invariance of the Baxter
equation under the substitution $ x \rightarrow -x
\,,\,\,\vec{\mu} \rightarrow \vec{\mu ^s} $.

The coefficients $C_r^{(t)}(\vec{\mu}) \; ,C_r^{(n-1-t)}(\vec{\mu^s})
$ and $ \beta ^{t}(\vec{\mu}) $ are obtained imposing the asymptotic
validity of the Baxter equation,
$$
\lim _{x \rightarrow \infty} \; x^k \; Q^{(t)}(x,\vec{\mu}) =0
\quad ,\quad k=1,2, \ldots ,n-2 \; .
$$
This leads to a system of $n-2$ linear equations
\bea
&&\sum_{r=1}^{t} G_{k,r}(\vec{\mu})\,C_r^{(t)}(\vec{\mu})+(-1)^k
\beta ^{(t)}(\vec{\mu})\,\sum
_{r=1}^{n-1-t}G_{k,r}(\vec{\mu^s})\,C_r^{(n-1-t)}(\vec{\mu^s})=0\,,
\cr \cr
&& k=1,2,\ldots  ,n-2 \; ,
\eea
where the matrix elements $G_{k,r}(\vec{\mu})$ are defined by
eq.(\ref{defG})
and we normalize $Q^{(t)}(x,\vec{\mu})$ by choosing
\bea
C_t^{(t)}(\vec{\mu})=C_{n-1-t}^{(n-1-t)}(\vec{\mu})=1 \; .
\eea
Moreover, one can  obtain the following relations using the symmetry of the
Baxter equation,
\begin{equation}\label{relsim}
Q^{(r)}(x,\vec{\mu}) =\beta ^{(r)}(\vec{\mu}) \;
Q^{(n-1-r)}(-x,\vec{\mu ^s}) \,,\,\,\beta ^{(r)}(\vec{\mu}) \; \beta
^{(n-1-r)}(\vec{\mu^s})=1 \, ,  \,\, \beta ^{(0)}(\vec{\mu})=1\,.
\end{equation}

It is important to notice that three subsequent solutions $Q^{(r)}$ for
$r=1,2,...,n-2$ are linearly related by
\begin{equation} \label{ecrecu}
\left[\delta ^{(r)} (\vec{\mu})+\pi \, \cot(\pi x)\right]
\,Q^{(r)}(x,\vec{\mu})=
Q^{(r+1)}(x,\vec{\mu})+\alpha ^{(r)}(\vec{\mu})\; Q^{(r-1)}(x,\vec{\mu})\,.
\end{equation}
Indeed, the left and right-hand sides satisfy the Baxter equation everywhere
including $x \rightarrow \infty$ and have the
same singularities. Due to the uniqueness of the `minimal'
solutions the quantity $\cot (\pi x)\,Q^{(r)}(x,\vec{\mu})$ can
be expressed as a linear combination of
$Q^{(r-1)}(x,\vec{\mu}),\,Q^{(r)}(x,\vec{\mu})$ and
$Q^{(r+1)}(x,\vec{\mu})$.
Furthermore, the coefficient in front of $Q^{(r+1)}(x,\vec{\mu})$ is
chosen to be $1$
taking into account our normalization of $Q^{(r)}(x,\vec{\mu})$.
In this normalization
we obtain
\begin{equation}
\alpha ^{(r)}(\vec{\mu}) =-\frac{\beta
^{(r)}(\vec{\mu})}{\beta ^{(r-1)}(\vec{\mu})}\,,\,\, \beta
^{(r)}(\vec{\mu}) =(-1)^r \prod _{t=1}^r \alpha
^{(t)}(\vec{\mu})\,.
\end{equation}
The following relations hold for the coefficients $\alpha
^{(r)}(\vec{\mu})$ and $\delta ^{(r)}(\vec{\mu})$ \begin{equation} \delta
^{(r)}(\vec{\mu})=-\delta ^{(n-1-r)}(\vec{\mu^s}) \quad , \quad
\alpha^{(r)}(\vec{\mu}) \; \alpha ^{(n-1-r)}(\vec{\mu^s})=1 \; .
\end{equation}

According to the above recurrence relations the functions
$Q^{(r)}(x,\vec{\mu})$
($r=1,2, \ldots ,n-2$) can
be expressed  as linear combinations of $Q^{(n-1)}(x,\vec{\mu})$
and $Q^{(0)}(x,\vec{\mu})$ with the coefficients being periodic functions
of $x$ \cite{hect}:
\begin{eqnarray}
&&D\left[ \vec{\delta} (\vec{\mu}),\,\vec{\alpha}(\vec{\mu})\,,
\pi \cot (\pi x) \right] \,Q^{(r)}(x,\vec{\mu})= \\ \cr
&& = D^{(r)}_0\left[ \vec{\delta}(\vec{\mu}) ,\,
\vec{\alpha}(\vec{\mu}),\pi \cot (\pi x) \right]
 \,Q^{(0)}(x,\vec{\mu})+ D^{(r)}_{n-1}\left[ \vec{\delta}
 (\vec{\mu}),\vec{\alpha}(\vec{\mu}), \pi \cot (\pi x)
 \right]\,Q^{(n-1)} (x,\vec{\mu})\,   \nonumber
\end{eqnarray}
compatible with the relation (\ref{relsim}).

The factor $D\left[ \vec{\delta} ,\,\vec{\alpha},\, \pi \cot
(\pi x )\right] $  can be written as the determinant of the matrix $\Lambda$
\begin{equation}
D\left[ \vec{\delta} ,\,\vec{\alpha},\, \pi \cot (\pi x )\right]=
\left| \left| \Lambda \left[\vec{\delta} ,\,\vec{\alpha},\, \pi
\cot (\pi x )\right]\right| \right|.
\end{equation}
The matrix
$\Lambda$ takes the form,
\begin{eqnarray}
\Lambda\left[\vec{\delta} ,\,\vec{\alpha},\, \pi \cot (\pi x
)\right]=\pi \, \cot (\pi x) \; I+  \Delta \left[\vec{\delta}
,\,\vec{\alpha}\right]\; , \; I= \left( \begin{array}{cccc} 1 & 0
& ...  & 0 \\ 0 & 1 & ... & 0 \\ ... & ...& ...& ...\\ 0 & 0 &
...  & 1 \end{array} \right)
\end{eqnarray}
and
\begin{eqnarray}
\Delta\left[ \vec{\delta} ,\,\vec{\alpha}\right] = \left(
\begin{array}{ccccc}
\delta ^{(1)}  & -1 & 0 & ...& 0 \\
- \alpha ^{(2)} & \delta ^{(2)}  & -1 & ...& 0\\
0 & -\alpha ^{(3)} & \delta ^{(3)}  & ...& 0\\
....&...&...&...&...\\
0 & 0 & 0 & ... & \delta ^{(n-2)}
\end{array}\right).
\end{eqnarray}
In an analogous way
$D^{(r)}_0\left[\vec{\delta} , \vec{\alpha},\pi \cot (\pi x)\right]$
and $D^{(r)}_{n-1} \left[\vec{\delta} , \vec{\alpha},\pi \cot (\pi
x)\right]$
can be expressed in terms of the  determinants of the matrices
$\Lambda ^{(r)}_0$ and $\Lambda ^{(r)}_{n-1}$ of (lower)
rank $n-3$ obtained from $\Lambda$ by removing the column $r$ and the
first or the last line, respectively:
\[
D^{(r)}_0\left[\vec{\delta} , \vec{\alpha},\pi \cot (\pi x)\right]= (-1)^r
\,
\alpha ^{(1)}  \,\left| \left| \Lambda ^{(r)}_0 \right| \right| \; , \;
D^{(r)}_{n-1}\left[\vec{\delta} , \vec{\alpha},\pi \cot (\pi x)\right]
= (-1)^r \; \left| \left| \Lambda ^{(r)}_{n-1} \right| \right|
\]
and
\[
\Lambda ^{(r)}_0=\left(
\begin{array}{ccccc}
- \alpha ^{(2)} & \delta ^{(2)}+\pi \cot (\pi x) & ... & ... &  0 \\
0 & - \alpha ^{(3)} & ... & ... & 0 \\
... & ... & \delta ^{(r+1)}+\pi \cot (\pi x) & ... & 0 \\
... & ... & ... & ... & ... \\
0 & 0 & ... & ... & \delta ^{(n-2)}+\pi \cot (\pi x)
\end{array}
\right),
\]
\[
\Lambda ^{(r)}_{n-1}=\left(
\begin{array}{ccccc}
\delta ^{(1)}+\pi \cot (\pi x) & -1 & ... & ... & 0 \\
- \alpha ^{(2)} & \delta ^{(2)}+\pi \cot (\pi x) & ... & ... &  0 \\
... & ... & -1 & ... & 0 \\
... & ... & ... & ... & ... \\
0 & 0 & ... & ... & -1
\end{array}
\right).
\]

Let us write these combinations in the form
\[
Q^{(r)}(x,\,\vec{\mu})=a^{(r)}[\vec{\mu} , \pi \cot (\pi x)] \;
Q^{(n)}[x,\,\vec{\mu}] 
+b^{(r)}[\vec{\mu} , \pi \cot (\pi x)] \,Q^{(0)}[x,\,\vec{\mu}]
\]
for $n+1$ reggeons and
\[
Q^{(r)}(x,\,\vec{\mu})=a'^{(r)}[\vec{\mu} , \pi \cot (\pi x)] \;
Q^{(n+1)}(x,\,\vec{\mu})
+b'^{(r)}[\vec{\mu} , \pi \cot (\pi x)] \; Q^{(0)}(x,\,\vec{\mu}) \; 
\]
for $n+2$ reggeons.
Then, the following relations are valid
\bea
&&a'^{(r)}[\vec{\mu} , \pi \cot (\pi x)]=a^{(r)}[\vec{\mu} , \pi \cot
(\pi x)] \;
a'^{(n)}[\vec{\mu} , \pi \cot (\pi x)]\; , \cr \cr
&&b'^{(r)}[\vec{\mu} , \pi \cot (\pi x)]=b^{(r)}[\vec{\mu} , \pi \cot (\pi x)]+
a^{(r)}[\vec{\mu} , \pi \cot (\pi x)] \; b'^{(n)}[\vec{\mu} , \pi \cot
(\pi x)] \; , 
\eea
where 
\bea
&&a'^{(n)}[\vec{\mu} , \pi \cot (\pi x)]=\frac{1}{
\delta ^{(n)}(\vec{\mu})+\pi \cot (\pi x)-\alpha ^{(n)}(\vec{\mu})
a^{(n-1)}(\vec{\mu})} \;
,\cr \cr
&&b'^{(n)}[\vec{\mu} , \pi \cot (\pi x)]=b^{(n-1
)}[\vec{\mu} , \pi \cot (\pi x)] \; \alpha^{(n)}( \vec{\mu}) \,
a'^{(n)}[\vec{\mu} , \pi \cot (\pi x)] \; .
\eea
Notice that  the linear relations (\ref{ecrecu}) among
$Q^{(r)}(x,\,\vec{\mu}),\,Q^{(r+1)}(x,\,\vec{\mu})$ and
$Q^{(r-1)}(x,\,\vec{\mu})$ are similar to the recurrence
relations for the orthogonal polynomials $P_r(z)$ if we substitute
$\pi \,\cot (\pi x)$ by a variable $z$. The Baxter
functions also belong to an orthonormalized set of functions.

We use below the formulae of this section for the numerical
calculations of the intercepts of the composite  states constructed from
reggeized gluons.

\section{Spectrum of eigenvalues of integrals of motion}

To quantize the integrals of motion one should impose all physical 
constraints on the corresponding eigenfunctions. In the impact 
parameter representation
$\vec{\rho}$ the wave function $\Psi $ should be normalized in a 
conformally - invariant way \cite{conf,integr} and the Hamiltonian 
$H$ in the space of
these functions should be hermitian. The integrals of motion $q_k$ 
($q_k^*$) are 
differential operators in the holomorphic (antiholomorphic) 
subspaces \cite{integr}. The wave function in the 
$\vec{\rho}$-representation is
constructed in a bilinear form from all eigenfunctions of the
integrals of motion similar to the case of two-dimensional 
conformal field theories in such a way that 
its single-valuedness holds. The single-valuedness condition is an
important constraint on the coefficients of the bilinear form and 
on the possible values of the integrals of motion. But there are other 
constraints which are significantly simpler.

In the case of the Pomeron one does not have extra integrals of motion
apart from two Casimir operators with their eigenvalues depending on
the conformal weights $m$ and $\widetilde{m}$. For the principal
series of the unitary representations
of the M\"{o}bius group the weights depend on the real parameter $\nu$
and the integer conformal spin $n$ \cite{conf}. 

For three particles there is an additional integral of motion with its
eigenvalue parametrized as $q_3=i\mu$. The problem is to find the
region to which the parameter $\mu$ belongs. For hermitian operators
the eigenvalues are real, but the integrals  of motion
$\hat{q_k},\hat{q_k^*}$ in holomorphic and anti-holomorphic subspaces
are not hermitian. With the use of conformal invariance the integral of motion 
$\hat{q_3}$ for the Odderon can be 
written as the third order differential operator \cite{intmot, dual}
\begin{equation}
\hat{q_3}=a_{1-m}\,a_{m}=z \, (1-z)\left[ z(1-z)p^3+i(2-m)(1-2z)p^2+
(2-m)(1-m)p\right] \; ,
\end{equation}
acting on the
functions depending on the anharmonic ratio
\begin{equation}
z=\frac{\rho _{12}\rho
_{30}}{\rho _{10}\rho _{32}}\,.
\end{equation}
Here $p=i\frac{\partial}{\partial z} $ and $a_m$ is the duality 
operator \cite{dual}
\begin{equation}
a_m=z(1-z) \,p^{1+m} \,,
\end{equation}
transforming a state with the conformal weight $m$ into a  state with
conformal weight $1-m$
$$
a_m \phi _m =l_m \phi _{1-m}\,,\,\,a_{1-m} \phi _{1-m} =l_{1-m} \phi 
_m \,.
$$    
Here $ l_m $ and $ l_{1-m} $ are the eigenvalues of the corresponding
transformations depending on the norms of the functions $\phi_m$. 
The product of these numbers does not depend on the normalization and 
is related with the eigenvalue $\mu$ by
$$
l_m \,l_{1-m}=q_3=i\mu \,.
$$

The Odderon wave function $\phi _{m, \widetilde{m}}(\vec{z})$ is the 
bilinear combination of 
holomorphic and anti-holomorphic functions \cite{JW}
\begin{equation}
\phi _{m, \widetilde{m}}(\vec{z})=\sum _{ik}
C_{ik}\left[ \phi ^i_{m,q_3}(z)\phi 
^k_{\widetilde{m},q_3^* }(z^*)+\phi ^i_{m,-q_3}(z)\phi
^k_{\widetilde{m},-q_3^* }(z^*)\right]\,,
\end{equation}
where the symmetrization under the transformation $q_3 \rightarrow -q_3$ is 
needed for the Bose symmetry of the wave function. The coefficients 
$C_{ik}$ are calculated from the single-valuedness of $\phi _{m, 
\widetilde{m}}(\vec{z})$. The sum 
is performed over three independent eigenfunctions of the integrals of 
motion $\hat{q}_3, \hat{q}_3^*$ in the holomorphic and anti-holomorphic
spaces. It was argued in ref.\cite{JW} that the eigenvalues  $q_3$ are
purely imaginary as a consequence of single-valuedness. We give below 
additional arguments supporting this conclusion [at least for real
values of $m(m-1)$]. 

To begin with, let us perform the simultaneous interchange of $(z, 
m)$ and $(z^*, \widetilde{m})$ in $\phi _{m,
\widetilde{m}}(\vec{z})$. The wave function should remain the same if
there is no 
accidental degeneracy. But this can be true only if $q_3$ satisfies one
of the two relations \cite{dual}
\begin{equation}
q_3=\pm q_3^* \,,
\end{equation} 
which is in agreement with the Janik-Wosiek result $q_3=- 
q_3^*$ \cite{JW}. An analogous argument for the general $n$-reggeons
case will lead to the guess that $\vec{\mu}^*=\vec{\mu}$ or
$\vec{\mu}^*=\vec{\mu}^s$.  

Furthermore, although the holomorphic hamiltonian $h$ is not a 
single-valued operator (see for example the 
discussion of the separability properties of the BFKL hamiltonian in 
ref.\cite{hect}), its 
ambiguity seems to be simple and related with the possibility to 
add to it some periodic functions of $\partial / \partial (\ln 
\rho _k)$ cancelling in the total hamiltonian $H$. For example, 
in the case of the Odderon one can write the holomorphic hamiltonian 
in the normal form \cite{dual} 
$$
h=-\ln (z)+\Psi (1-P)+\Psi (-P)+\Psi (m-P)-3 \Psi 
(1)+\sum 
_{k=1}^{\infty}z^k\,f_k(P) \,, 
$$
where $z$ is the anharmonic ratio, $ P= \partial / \partial (\ln z ) $ and 
the explicit formulae for $f_k (P)$ are given in ref.\cite{dual},
where $h$ was defined with an extra factor $2$  
in comparison with the present paper and with ref.\cite{hect}.  The above
expression for $h$ is equivalent to the expressions obtained by
adding to it one of two terms
$$
-2\pi \cot (\pi P)\,,\,\,\pi \left[\cot (\pi (m-P))-\cot (\pi P)\right]\,, 
$$ 
cancelling poles of  the $\Psi$ -functions. Apart from such 
ambiguities, $h$ can be considered for real $m(m-1)$ as a symmetric operator 
$h=h^T$ acting on functions $\phi (z, z^*)$ depending on real
variables $z, z^*$ with the norm
$$
\int _{-\infty}^{+\infty}\frac{dz^*}{z^*(1-z^*)} 
\int _{-\infty}^{+\infty}\frac{dz}{z(1-z)} \; |\phi |^2\; .
$$
The integration over the real values of $z$ and $z^*$ can be obtained 
after the
anti-Wick rotation of the vector $\vec{z}$ to the Minkowski space.
Therefore the eigenvalues $\epsilon$ of $h$ should be real 
(for the case when $m(m-1)$ is real). On the other hand, 
$\epsilon$ is a real function of the eigenvalue $\mu ^2$
(apart from constant imaginary contributions cancelling
in the total energy $E$). 
For example,  we have for large $\mu$ the expansion of
$\epsilon$ in $\mu ^{-2}$ \cite{dual},
$$
\epsilon=\ln (\mu )+3 \gamma +\left[
\frac{3}{448}+\frac{13}{120}(m-1/2)^2-\frac{1}{12}(m-1/2)^4\right]
\frac{1}{\mu ^2}+{\cal O}\left(\frac{1}{\mu ^4} \right) \, .
$$
Therefore $\mu$ is real or pure imaginary for real $m(m-1)$ (at least 
for large $|\mu |$). 

Because the operator $q_3$ is anti-symmetric 
under the permutation of the coordinates $\rho _k$ and $\rho _l$, its matrix 
element between the symmetric wave functions is zero, but the matrix
element of $q_3^2$ can be written as follows
$$
\int _{-\infty}^{+\infty}\frac{dz^*}{z^*(1-z^*)}
\int _{-\infty}^{+\infty}\frac{dz}{z(1-z)} \,\hat{q}_3\phi
\,\hat{q}_3\phi ^*
\;\; 
$$
and therefore the eigenvalues $\mu$ of $-i\hat{q}_3$ are again real or pure 
imaginary for real $m(m-1)$.

Let us consider now the expression for the total energy of the composite 
state of $n$-reggeons in terms of the 
holomorphic and anti-holomorphic energies
$$
E_{m, \widetilde{m}}=\epsilon _{m}(\vec{\mu})+
\epsilon _{\widetilde{m}}(\vec{\mu} ^{s*})
$$
valid for the wave function $ \phi _{m,\widetilde{m}} $ satisfying the 
Schr\"{o}dinger equation in the Baxter-Sklyanin 
representation in the limit $\lambda , \lambda ^* \rightarrow 
i$ \cite{hect}.
We can obtain the analogous expression 
$$
E_{m, \widetilde{m}}=\epsilon _m(\vec{\mu}^s)+\epsilon _{\widetilde{m}}
(\vec{\mu} ^*)
$$
by taking instead another limit $\lambda ,\lambda ^* \rightarrow 
-i$.
These two expressions for energies were derived from the Schr\"{o}dinger
equation with the
hermitian hamiltonian \cite{hect}. Therefore they should coincide for its
eigenfunction $\phi 
_{m,\widetilde{m}}$. This is possible only if the following 
property is fulfilled for the quantized values of $\vec{\mu}$:
$$
\epsilon _m (\vec{\mu})+\epsilon _{\widetilde{m}}(\vec{\mu }
^{s*})=\epsilon _{m}
(\vec{\mu }^s)+\epsilon _{\widetilde{m}}(\vec{\mu} ^*) \,,
$$
The solution of this equation for real $m(m-1)$ 
is 
\be \label{condmu}
\vec{\mu} ^*=\vec{\mu} ,
\ee
or $\vec{\mu}^*=\vec{\mu}^s$, if again there is no accidental degeneracy.

The arguments given in this section support our assumption  \cite{hect},
that the eigenvalues $\vec{\mu}$ are real for real $m(m-1)$. For non-vanishing
conformal spins $n$ the values of $\epsilon$ and $\vec{\mu}$ can be obtained
by analytic continuation of the anomalous dimensions of the
corresponding high-twist operators with $n=0$ to continuous values of
the Lorentz spin $j$ (see sections 6-8 below). 

In our opinion the complex values
for $\mu$ found in \cite{num} do not correspond to eigenfunctions
of the Schr\"{o}dinger equation for the Odderon and they
were obtained due to an incorrect quantization procedure. 

\section{Zeroes of the Baxter function and the quantization of the
integrals of motion}

The zeroes of the Baxter function are very important, because their
position $\lambda _k$ is fixed by the Bethe equations and their
knowledge gives
a possibility to write the wave function of the composite states in the
framework of the Bethe Ansatz as an (infinite) product of the
differential operators  $B(\lambda _k)$ applied to the pseudo-vacuum
state. The positions of the zeroes of $Q(\lambda)$ for the Pomeron wave
function was investigated in our previous paper \cite{hect}. For the case
of the composite states constructed from $n>2$ reggeons the number of the
`minimal' solutions $Q^{(r)}$ of the Baxter equation is  $n$ and their
linear
combinations have rather complicated sets of zeroes.

However, certain linear combinations of functions $Q^{(r)}(x,\,\vec{\mu})$
have zeroes which are situated at equidistant points $x_k=x_0+k, \;
k \in {\cal Z} $.
Indeed, according to  relations (\ref{ecrecu})
among the Baxter functions
$Q^{(r)}(x,\,\vec{\mu}), \; Q^{(r+1)}(x,\,\vec{\mu})$
and $Q^{(r-1)}(x,\,\vec{\mu})$ such situation happens for
solutions of the equation 
\be \label{eccero}
Q^{(r+1)}(x,\vec{\mu})+\alpha
^{(r)}(\vec{\mu})\,Q^{(r-1)}(x,\vec{\mu})=0\,,
\ee
where $r=1,2, \ldots ,n-2$.
Some of the solutions of eq.(\ref{eccero}) coincide with the zeroes of
the function   $Q^{(r)}(x,\,\vec{\mu})$ while other roots are
situated at the equidistant points
\be \label{ceros}
x_k^{(r)}(\vec{\mu})=k -\frac{1}{\pi}\,\mbox{arccot}\left[ \,\frac{\delta
^{(r)}(\vec{\mu})}{\pi}\right]\,,\,\, \; k \in {\cal Z}.
\ee

For the eigenstates of the Hamiltonian the parameters
$\vec{\mu}$ are quantized in accordance with the physical
requirement \cite{hect} that the holomorphic energies for
different Baxter functions are the same. It is equivalent to
the condition, that all
parameters $\delta^{(r)}(\vec{\mu})$ vanish
\begin{equation}
\delta ^{(r)}(\vec{\mu})=0 \; ,
\end{equation}
because otherwise, the
energies for the solutions $Q^{(r)}(x,\vec{\mu})$ and
$Q^{(r+1)}(x,\vec{\mu})$ would not coincide. Eq.(\ref{ceros})
then implies that for quantized $\vec{\mu}$ the above linear
combination of $Q^{(r+1)}(x,\vec{\mu})$ and
$Q^{(r-1)}(x,\vec{\mu})$ has a sequence of zeroes at the points
$$
x_k=k+\frac{1}{2}\,,\, \; k \in {\cal Z}.
$$

Let us consider now the superpositions of the Baxter
functions $Q^{(n-1)}(x,\vec{\mu})$ and $Q^{(0)}(x,\vec{\mu})$ with
their poles situated only at positive and negative integer
points, respectively.
It is obvious from the previous section that there are $n-2$
different linear combinations of these functions
\begin{equation}
\Phi^{(t)}(x,\vec{\mu})
=Q^{(n-1)}(x,\vec{\mu})+c^{(t)}(\vec{\mu})\,Q^{(0)}(x,\vec{\mu})
\end{equation}
having equidistant zeroes at the points $x=x_k$ where (see (58))
\begin{equation}
D\left[\vec{\delta} (\vec{\mu}),\,\vec{\alpha}(\vec{\mu})\,,
\pi \cot (\pi x)\right]=0 \; .
\end{equation}
The last equation has $n-2$ different solutions
\begin{equation}
z^{(t)}(\vec{\mu})=\pi \cot \left[ \pi x^{(t)}(\vec{\mu})\right]
\,,\,\,t=1,2,...n-2\,
\end{equation}
and for each solution there is a linear combination of the Baxter
functions $Q^{(n-1)}(x,\vec{\mu})$ and $Q^{(0)}(x,\vec{\mu})$ with
the relative coefficient
\begin{equation}
c^{(t)}(\vec{\mu})=\frac{D^{(r)}_{0}\left[ \vec{\delta}
(\vec{\mu}),\vec{\alpha}(\vec{\mu}) \; ,
z^{(t)}(\vec{\mu}) \right] }{D^{(r)}_{n-1} \left[  \vec{\delta}
(\vec{\mu}),\vec{\alpha}(\vec{\mu}), z^{(t)}(\vec{\mu})\right] }\, .
\end{equation}
Note, that $c^{(t)}(\vec{\mu})$ does not depend on the parameter $r$, if
there is no accidental degeneracy.

In the case, when
\begin{equation}
\delta ^{(r)}(\vec{\mu})=0
\end{equation}
for all $r=1,2,...,n-2$, which corresponds to the quantization of the
integrals of motion $\vec{\mu}$, the determinant $D\left[\vec{\delta}
(\vec{\mu}),\,\vec{\alpha}(\vec{\mu}),\,\pi \cot (\pi x)\right] $ is an even
(odd) function of its
argument $\pi \cot (\pi x)$  for even  (odd) $n$. Therefore,
the equidistant zeroes of two different functions
$\Phi^{(t)}(x,\vec{\mu})$  (or zeroes of the same function
$\Phi^{(t)}(x,\vec{\mu})$) have opposite signs (modulo an
integer number):  \begin{equation}
x^{(t)}(\vec{\mu})=-x^{(n-2-t)}(\vec{\mu}) \; .
\end{equation}
For odd $n$ one function  has a sequence of zeroes at
$$
x_{k}=k+\frac{1}{2}\,.
$$

In a general case of  $n$ reggeized gluons we can calculate
the important coefficients $\alpha ^{(r)}(\vec{\mu})$ in the 
recurrent relations and
the quantized values of $\vec{\mu}$ only by finding $Q^{(n)}$,
$Q^{(0)}$ and all their linear combinations with constant
coefficients which have the equidistant zeroes with above
properties. Let us consider several examples:

For $n=3$ we have only one function  with equidistant zeroes $x_k=k+1/2$
$$
Q(x)=Q^{(2)}(x,\mu)+\alpha^{(1)}(\mu ) \; Q^{(0)}(x,\mu)\,.
$$
For $n=4$ there are two functions,
$$
Q_{\pm}=Q^{(3)} \pm \alpha ^{(1)} \sqrt{\alpha ^{(2)}}\,Q^{(0)}
$$
with zeroes at $ \pi \cot {\pi x} =\pm \sqrt{\alpha _2} $,
respectively.
For $n=5$ there are also two functions with equidistant
zeroes:
\[
Q_1=Q^{(4)} + \alpha ^{(1)}\,\alpha ^{(2)} \,Q^{(0)}
\,,\,\,
Q_2=Q^{(4)} - \alpha ^{(1)}\,\alpha ^{(3)}\,Q^{(0)} 
\] 
with the equidistant zeroes at $\pi \cot {\pi x}=\pm 
\sqrt{\alpha ^{(2)}+\alpha ^{(3)}}$ and $\pi \cot {\pi x}=0$, respectively.
In all these cases, one can calculate all $\alpha ^{(r)}$ and the
quantized values of $\vec{\mu}$ by finding  the corresponding
combinations of $Q^{(n)}$ and $Q^{(0)}$ with the equidistant zeroes.

 \section{Anomalous
dimensions of quasi-partonic operators}

The $Q^{2}$-dependence of the inclusive probabilities $n_{i}(x,\ln Q^{2})$
to have a parton $i$ with momentum fraction $x$ inside a
hadron with large momentum $\left| \overrightarrow{p}\right|
\rightarrow \infty $
can be found from the DGLAP evolution equation \cite{DGLAP}. The eigenvalues
of its integral kernels describing the inclusive parton transitions
$i\rightarrow k$ coincide with the matrix elements
$\gamma _{j}^{ki}(\alpha )$
of the anomalous dimension matrix for twist-2 operators $O^{j}$ with the
Lorentz spins $j=2,3,...\,$. These operators are bilinear in the
gluon ($i=g$) or quark ($i=q$) fields. For example, the twist-2
gluon operator with the Lorentz spin $j$ can be  written as,
\begin{equation}\label{oper}
O_{.....}^{j}=n^{\mu _{1}}n^{\mu _{2}}...n^{\mu _{j}}\,tr\,\,G_{\rho \mu
_{1}}D_{\mu _{2}}D_{\mu _{3}}...D_{\mu _{j-1}}G_{\rho \mu _{j}}\,,
\end{equation}
where $D_{\mu }=\partial _{\mu }+g\,V_{\mu }$, $V_{\mu }$ 
and $G_{\rho \mu}=\frac{1}{g}\left[ D_{\rho }\,,D_{\mu }\right] $ 
are the covariant derivative, the gluon field and the 
field tensor, respectively.

The symmetric 
traceless tensor $O_{\mu_{1}\mu _{2}...\mu _{j}}$ is multiplied by
the light-cone vectors $n_{\mu_{r}}$
\[
n_{\mu }=q_{\mu }+x\,p_{\mu }\,,\,\,n_{\mu }^{2}=0\,,\,\,p_{\mu }^{2}\simeq
0\,,\,\,q_{\mu }^{2}=-Q^{2},\,\,x=\frac{Q^{2}}{2pq}\,,
\]
where $p_{\mu}$ and $q_{\mu}$ in the deep-inelastic
$ep$ scattering are the momenta of the
initial proton and virtual photon, respectively.

The matrix elements of the operators $O_{.....}^{j}$ between the hadron
states are renormalized as functions of the growing ultraviolet cut-off
$Q^{2}$. For example, in the case of the pure Yang-Mills theory with the
gauge group $SU(N_{c})$ we have

\[
\left\langle p\left| O_{.....}^{j}\right| p\right\rangle \sim \exp \left(
\int_{Q_{0}^{2}}^{Q^{2}}\gamma _{j}(\alpha _{s}(Q^{\prime 2}))\,d\ln
Q^{\prime 2}\right) \,,\,\,\,\alpha _{s}(Q^{2})\simeq \frac{4\pi }{\beta
_{2}\ln \frac{Q^{2}}{\Lambda _{s}^{2}}}\,,\,\,\,\beta _{2}=\frac{11}{3}N_{c}
- \frac{2}{3}n_{f}\,.
\]
where $\Lambda _{s}\simeq 200\,Mev$ is the QCD parameter and the anomalous
dimension $\gamma _{j}(\alpha )$ can be calculated perturbatively as
\begin{equation}
\gamma _{j}(\alpha )=\sum_{k=1}^{\infty }C_{j}^{(k)}\left( \frac{\alpha
N_{c}}{\pi }\right) ^{k}\,.
\end{equation}
In particular, to the lowest order
\[
C_{j}^{(1)}=\Psi (1)-\Psi (j-1)-\frac{2}{j}+\frac{1}{j+1}-\frac{1}{j+2}
+\frac{11}{12}\,.
\]
The gluon anomalous dimension is singular in the non-physical point $\omega
=j-1\rightarrow 0$. In this limit one can calculate it to all orders of
perturbation theory \cite{conf}
\begin{equation}\label{gamaL}
\gamma _{\omega }=\frac{\alpha N_{c}}{\pi \omega }
- \Psi ^{\prime \prime }(1)\,\left( \frac{\alpha N_{c}}{\pi
\omega }\right) ^{4}+...
\end{equation}
from the eigenvalue of the kernel of the BFKL equation in LLA \cite{BFKL}
at $n=0$:
\begin{equation}
\omega _{BFKL}=\frac{\alpha \,N_{c}}{\pi }\left[ 2\Psi (1)-\Psi (\gamma
)-\Psi (1-\gamma )\right] \,.
\end{equation}
Notice that the coefficients of $ \left( \frac{\alpha
N_{c}}{\pi \omega }\right) ^{2} $ and $ \left( \frac{\alpha N_{c}}{\pi
\omega }
\right) ^{3} $ {\bf exactly } vanish in eq.(\ref{gamaL}). Here $ \Psi
^{\prime \prime }(1) = - 2 \zeta(3) = - 2.4041138 \ldots $.

Indeed, the Green function satisfying the inhomogeneous BFKL equation can be
written as follows \cite{conf}
\begin{equation}
<\,\phi (\overrightarrow{\rho _{1}})\,\phi (\overrightarrow{\rho _{2}}
)\,\phi (\overrightarrow{\rho _{1^{\prime }}})\,\phi (\overrightarrow{\rho
_{2^{\prime }}})\,>\,=\sum_{n}\int_{-\infty }^{+\infty }d\nu \; C(\nu ,n)\int
d^{2}\rho _{0}\,\frac{E_{\nu ,n}(\overrightarrow{\rho _{10}},
\overrightarrow{\rho _{20}})\,E_{\nu ,n}^{\ast }
(\overrightarrow{\rho _{1^{\prime }0}},
\overrightarrow{\rho _{2^{\prime }0}})}{\omega -\omega ^{0}(n,\nu )}\,,
\end{equation}
where $\omega ^{0}(n,\nu )$ is the eigenvalue 
of the BFKL kernel and $C(\nu ,n)$ is fixed by
the completeness condition for its eigenfunctions
$$
E_{\nu ,n}(\overrightarrow{\rho _{10}},\overrightarrow{\rho _{20}})=
<0|\phi (\vec{\rho}_1) \,\phi (\vec{\rho}_2)\, O_{\nu ,n}
(\vec{\rho }_0)|0>=
$$
\be \label{pesos}
\left(
\frac{\rho _{12}}{\rho _{10}\rho _{20}}\right) ^{m}\left( \frac{\rho
_{12}^{\ast }}{\rho _{10}^{\ast }\rho _{20}^{\ast }}\right) ^
{\widetilde{m}}
\,,\,\,m=\frac{1}{2}+i\nu
+\frac{n}{2}\,,\,\,\widetilde{m}=\frac{1}{2}+i\nu
- \frac{n}{2}\,.
\ee

In the limit $\left| \rho _{1^{\prime }2^{\prime }}\right| \rightarrow 0$
one can perform a Wilson expansion for the product of the fields $\phi
(\overrightarrow{\rho _{1^{\prime }}})$ and $\,\phi (\overrightarrow{\rho
_{2^{\prime }}})$. In this case the integral over
$\overrightarrow{\rho _{0}}$ can be calculated by extracting the factor
$E_{\nu ,n}(\overrightarrow{\rho _{10}},\overrightarrow{\rho _{20}})$ from
the integral at the point $\rho _{0}=\rho _{1^{\prime }}$ and by using the
relation
\[
\int d^{2}\rho _{0}\,E_{\nu ,n}^{\ast }
(\overrightarrow{\rho _{1^{\prime }0}}
,\overrightarrow{\rho _{2^{\prime }0}})\sim \rho _{1^{\prime }2^{\prime
}}^{m}\,\rho _{1^{\prime }2^{\prime }}^{\ast \widetilde{m}}\,\sim \left(
\frac{\rho _{1^{\prime }2^{\prime }}}{\rho _{1^{\prime }2^{\prime }}^
{\ast }}
\right) ^{\frac{n}{2}}\,\left| \rho _{1^{\prime }2^{\prime }}\right|
^{2\Gamma },\,\,\,\Gamma =\frac{m+\widetilde{m}}{2}=1-\gamma \,.
\]
Moreover at $\left| \rho _{1^{\prime }2^{\prime
}}\right| \rightarrow 0$ one can then shift the integration contour 
into the lower half of the $\nu $-plane up to the first pole of $[\omega
-\omega ^{0}(n,\nu )]^{-1}$  having $Im \,\nu<0 $ \cite{conf,report}:
\[
\lim_{\left| \rho _{1^{\prime }2^{\prime }}\right| \rightarrow 0}<\,\phi
(\overrightarrow{\rho _{1}})\,\phi (\overrightarrow{\rho _{2}})\,\phi
(\overrightarrow{\rho _{1^{\prime }}})\,\phi (\overrightarrow{\rho
_{2^{\prime }}})\,>\sim
\]
\begin{equation}
\sum_{n}e^{i\varphi n}\int_{\varepsilon -i\infty }^{\varepsilon +i\infty }
\frac{\widetilde{C}(\gamma ,n)\,\,d\,\gamma }{\omega -\omega ^{0}(n,\nu )}
\,\,\left| \frac{\rho _{12}\,\rho _{1^{\prime }2^{\prime }}}{\rho
_{11^{\prime }}\rho _{22^{\prime }}}\right| ^{2(1-\gamma )},\,\,\,\gamma =
\frac{1}{2}-i\nu \,,
\end{equation}
where $\varepsilon \sim \alpha _s / \omega$ and
\[
e^{i\varphi }=\sqrt{\frac{\rho _{11^{\prime }}^{\ast }\rho _{22^{\prime
}}^{\ast }\rho _{12}\rho _{1^{\prime }2^{\prime }}}{\rho _{12}^{\ast }\rho
_{1^{\prime }2^{\prime }}^{\ast }\rho _{11^{\prime }}\rho _{22^{\prime }}}}
\,.
\]
Note, that in accordance with the fact that the Green function includes the
external gluon propagators, the
scattering amplitude behaves in the momentum space as $\left|
k\right| ^{-4+2\gamma }$ for large gluon virtualities $\left| k\right| ^{2}$.

In particular, for $n=0$ and $\gamma \rightarrow 0$ one can obtain the above
expansion (\ref{gamaL}) in powers of $\frac{\alpha }{\pi \omega }$ for the
position $\gamma _{\omega 
}$ of the pole $(\gamma -\gamma _{\omega })^{-1}$. It is important that for
real $\omega $ the pole is situated on the real axis.
Therefore, the description of the corresponding state in the
framework of the M\"{o}bius group approach requires the
exceptional series of unitary representations \cite{Zhel} (contrary to the
Regge kinematics, where the principal series is used). For the
exceptional series the anomalous dimension
\[
\gamma =1-\frac{m+\widetilde{m}}{2}
\]
is real.

One can calculate from the BFKL equation also
anomalous dimensions of higher twist operators by solving the
eigenvalue equation near other singular points $\gamma =-k$
($k=1,2,...$) or by including in it a dependence from the
conformal spin $|n|$, which also leads to a shift of the pole
position to $\gamma =-|n|/2$ (see \cite{KoLi}).

But a more important problem is the calculation of the anomalous dimensions
for the so-called quasi-partonic operators  \cite{B'F'KL} constructed
from several gluonic fields. Indeed, the contribution of these operators at
$j\rightarrow 1$ is responsible for the unitarization of
structure functions at high energies. The simplest operator of
such type is the product of the twist-2 gluon operators

\begin{equation}
O^{j}=\prod_{r=1}^{p}O_{....}^{j_{r}}\,,\,\,\,\,j=\sum_{r=1}^{p}j_{r}=p+
\omega \,,\,\,\omega =\sum_{r=1}^{p}\omega _{r}\,.
\end{equation}

In the limit $N_{c}\rightarrow \infty $ this operator is multiplicatively
renormalized \cite{levin} and its dimension is the sum of dimensions of its
factors (including their anomalous dimensions $\gamma (\omega _{r})$)

\begin{equation}
\Gamma =p\,-\gamma \,,\,\,\gamma =\sum_{r=1}^{p}\gamma (\omega _{r}) \,,
\end{equation}
where in the expression for
the total dimension $\Gamma $ we neglected the small contribution
$\frac{\omega }{2}$.

To investigate the multi-Pomeron configuration, let us write the
hamiltonian describing the corresponding composite state in LLA
as a sum of pairwise BFKL hamiltonians

\[
H=\sum_{r=1}^{p}H_{a_{r}b_{r}}\,,
\]
neglecting the relative Pomeron interactions. In this case
its eigenfunction as a product of the Pomeron wave functions:

\[
\Psi (\rho _{a_{1}},\rho _{b_{1}},\rho _{0_{1}};...;\rho _{a_{p}},\rho
_{b_{p}},\rho _{0_{p}})=\prod_{r=1}^{p}\left( \frac{\rho _{a_{r}0_{r}}\rho
_{b_{r}0_{r}}}{\rho _{a_{r}b_{r}}}\right) ^{m_{r}}\left( \frac{\rho
_{a_{r}0_{r}}^{\ast }\rho _{b_{r}0_{r}}^{\ast }}{\rho _{a_{r}b_{r}}^{\ast }}
\right) ^{\widetilde{m}_{r}},
\]
where $\rho _{a_{r}},\,\rho _{b_{r}}$ and $\rho _{0_{r}}$ are the
coordinates of gluons $a_{r},\,b_{r}$ and  Pomerons $0_{r}$,
respectively, and the quantities $m_{r}$ and $\widetilde{m}_{r}$ are their
conformal weights. The wave function $\Psi $ belongs to a reducible
representation of the M\"{o}bius group and can be expanded in a sum
of irreducible representations with the use of the
Clebsch-Gordon coefficients
$C_{m_{1}\widetilde{m}_{1};...;m_{r}\widetilde{m}_{r}}^{m,\widetilde{m}}$
\cite{Zhel}

\[
\prod_{r=1}^{p}O_{m_{r}\widetilde{m}_{r}}(\overrightarrow{\rho }
_{0_{r}})=\sum_{m,\widetilde{m}}\int d^{2}\rho _{0}\,C_{m_{1}\widetilde{m}
_{1};...;m_{r}\widetilde{m}_{r}}^{m,\widetilde{m}}(\overrightarrow{\rho }
_{0_{1}},\,\overrightarrow{\rho }_{0_{2}},...,\,\overrightarrow{\rho }
_{0_{p}};\,\overrightarrow{\rho }_{0})\,\,\Phi ^{m,\widetilde{m}}
(\overrightarrow{\rho _{0}}).
\]
The contribution to the scattering amplitude in the coordinate space from
each irreducible component can be written as follows
\[
\prod_{r=1}^{p}\left[ \sum_{n_{r}=0}^{\infty }\int_{-\infty }^{\infty }d\nu
_{r}\int d^{2}\rho _{0_{r}}\left( \frac{\rho _{a_{r}b_{r}}}{\rho
_{a_{r}0_{r}}\rho _{b_{r}0_{r}}}\right) ^{m_{r}}\,\left( \frac{\rho
_{a_{r}b_{r}}^{\ast }}{\rho _{b_{r}0_{r}}^{\ast }\rho _{a_{r}0_{r}}^{\ast }}
\right) ^{\widetilde{m}_{r}}\right] \,\frac{C_{m_{1}\widetilde{m}
_{1};...;m_{r}\widetilde{m}_{r}}^{m,\widetilde{m}}(\overrightarrow{\rho }
_{0_{1}},\,...,\,\overrightarrow{\rho }
_{0_{p}};\,\overrightarrow{\rho }_{0})}{\omega -\sum_{s=1}^{p}\omega
(n_{s},\nu _{s})}\,.
\]

In the Regge regime $m_{r}$ and $\widetilde{m}_{r}$ belong to the principal
series of the unitary representations of the M\"{o}bious group
and therefore $m$ and $\widetilde{m}$ also belong to the same
series \cite{Zhel}. It means, that the position $\omega _{0}$ of
the $t$-channel partial wave singularity related to the
asymptotics of the cross-section $\sigma _{t}\sim s^{\omega
_{0}}$ equals $\omega _{0}=p\,\omega _{BFKL}$.

In the deep-inelastic regime the essential intervals are
small $\left| \rho _{00_{r}}\right| \sim 1/Q\ll \left| \rho
_{a_{r}b_{t}}\right| $. From dimensional considerations we
obtain at large $|Q|$ after integration over the essential
region of $\rho _{0_{r}}$ a power asymptotics of the
scattering amplitude

\begin{equation}
A^{(m,\widetilde{m})}\sim \prod_{r=1}^{p}\left[ \sum_{n_{r}=0}^{\infty
}\int_{\infty }^{\infty }d\nu _{r}\,\left( \frac{\rho _{a_{r}b_{r}}}{\rho
_{a_{r}0}\rho _{b_{r}0}}\right) ^{m_{r}}\,\left( \frac{\rho
_{a_{r}b_{r}}^{\ast }}{\rho _{b_{r}0}^{\ast }\rho _{a_{r}0}^{\ast }}\right)
^{\widetilde{m}_{r}}\right] \frac{C_{m_{1}\widetilde{m}_{1};...;m_{p}
\widetilde{m}_{p}}^{m,\widetilde{m}}\left| Q\right| ^{-2(p-\gamma
)}e^{in\varphi }}{\omega -\sum_{s=1}^{p}\omega (n_{s},\nu _{s})}\,,
\end{equation}
where $e^{i\varphi }=\sqrt{Q/Q^{\ast }}$, $n=\frac{1}{2}\sum_{r=1}^{p}(m_{r}
- \widetilde{m}_{r})$ and

\begin{equation}
\gamma =p-\frac{1}{2}\sum_{r=1}^{p}(m_{r}+\widetilde{m}_{r})
\end{equation}
is the anomalous dimension of a composite operator which turns out to be
real and small for small $g^{2}/\omega $. This expression 
for $\gamma$ is in
agreement with the known result \cite{Zhel}, that in the product of the
exceptional representations with real $\gamma _{r}=1-\frac{m_{r}+
\widetilde{m}_{r}}{2}>0$
there is a continuous spectrum of the unitary representations of the
principal series and only one representation from the
exceptional series having
\begin{equation}
\gamma
=\sum_{r=1}^{p}\gamma _{r}\,.
\end{equation}
Since $\omega (n,\nu )$ has a negative second derivative $\omega _{\gamma
\gamma }^{\prime \prime }(n,\nu )$ at $0<\gamma <1$, we obtain
after the integration over $\nu _{r}$ with the use of the
saddle-point method the following result for $A$ in the Bjorken regime
\begin{equation}
A^{(m,\widetilde{m})}\sim \prod_{r=1}^{p}\left[ \left( \frac{\rho
_{a_{r}b_{r}}}{\rho _{a_{r}0}\rho _{b_{r}0}}\right) ^{\frac{m}{p}}\,\left(
\frac{\rho _{a_{r}b_{r}}^{\ast }}{\rho _{b_{r}0}^{\ast }\rho _{a_{r}0}^{\ast
}}\right) ^{\frac{\widetilde{m}}{p}}\right] \,\left| Q\right| ^{-2(p-\gamma
)}\,,
\end{equation}
where
\begin{equation}
\gamma =p \; \omega ^{(-1)}(\omega /p)\,.
\end{equation}
and $\omega ^{(-1)}(\omega )$ is the inverse function to $\omega =\omega
_{BFKL}(\gamma )$.

In particular, for very large $Q^{2}$ corresponding to $\gamma \rightarrow
0$ and $n=0$ we have \cite{levin}
\begin{equation}
\gamma = \frac{\alpha _{s}N_{c}}{\pi \omega }\; p^{2} \; .
\end{equation}
It is possible to calculate also a correction of the relative
order $N_{c}^{-2}$ to this expression \cite{levin}.

In accordance with the topics of this paper let us consider now the high 
energy asymptotics of
the irreducible Feynman diagrams in which each of $n$ reggeized gluons
at $N_c \rightarrow \infty$
interacts only with two neighbours. In the Born approximation the
corresponding Green function is the product of
free gluon Green functions $\prod_{r=1}^{n}\ln \left| \rho
_{r}-\rho _{r}^{\prime }\right| ^{2}$. In LLA
$\frac{\alpha _{s}}{\omega}\sim 1$ it can be written as follows
(cf. \cite{conf})
\[
<\,\phi (\overrightarrow{\rho _{1}})...
\phi (\overrightarrow{\rho _{n}})
\,\phi (\overrightarrow{\rho _{1}^{\prime }})...\phi (\overrightarrow{\rho
_{n}^{\prime }})\,>=
\]
\begin{equation}
\sum_{n}\int_{-\infty }^{+\infty }d\nu \sum_{\mu _{3}}
...\sum_{\mu _{n}}\,C_{m,\widetilde{m};\mu _{3},...}\int d^{2}\rho
_{0}\,\frac{\,\Psi _{m,\widetilde{m};\mu _{3},...}(\overrightarrow{\rho }
_{1},..., \overrightarrow{\rho} _n;\,\overrightarrow{\rho
}_{0})\,\Psi _{m,\widetilde{m};\mu _{3},...}^{\ast
}(\overrightarrow{\rho }_{1}^{\prime },..., \overrightarrow{\rho}
_n^{\prime};\, \overrightarrow{ \rho }_{0}^{\prime })}{\omega
- \omega (m,\widetilde{m};\mu _{3},...,\mu _{n})}\,,
\end{equation}
where
$\omega (m,\widetilde{m};\mu _{3},...,\mu _{n}) \sim
\alpha _s$ are the eigenvalues of $H$ depending on the quantized
integrals of motion $\mu _{3},...,\mu _{n}$ and $\Psi
_{m,\widetilde{m};\mu _{3},...,\mu _{n}}(\overrightarrow{\rho
}_{1},\,\overrightarrow{\rho }_{2},...,\, \overrightarrow{\rho
}_{n};\,\overrightarrow{\rho }_{0})$ are the
corresponding eigenfunctions. From the M\"{o}bius
invariance of the Schr\"odinger equation we obtain
\begin{equation}
\Psi _{m,\widetilde{m};\mu _{3},\mu _{2},...,\mu _{n}}
(\overrightarrow{\rho }
_{1},...,\,\overrightarrow{\rho }_{n};\,\overrightarrow{\rho }_{0})\sim
\left( \frac{\rho _{12}\rho _{23}...\rho _{n1}}{\rho _{10}^{2}\rho
_{20}^{2}...\rho _{n0}^{2}}\right) ^{\frac{m}{n}}\left( \frac{\rho
_{12}^{\ast }\rho _{23}^{\ast }...\rho _{n1}^{\ast }}{\rho _{10}^{\ast
2}\rho _{20}^{\ast 2}...\rho _{n0}^{\ast 2}}\right)^
{\frac{\widetilde{m}}{n}}
f_{m,\widetilde{m}}(\overrightarrow{x}_{1},...,\overrightarrow{x}_{n-2}
)\,,
\end{equation}
where $\overrightarrow{x}_{r}$ ($x_{r}$ and $x_{r}^{\ast }$) are independent
anharmonic ratios of $\rho _{k}\,$\ and $\rho _{k}^{\ast }$ for
$k=0,1,...,n$.

In the Bjorken region, where
\[
\left| \rho _{r}^{\prime }-\rho _{s}^{\prime }\right| \sim Q^{-1}\ll \left|
\rho _{r}-\rho _{s}\right|\,,
\]
the essential integration domain is $\left| \rho _{0}-\rho _{r}^{\prime 
}\right| \sim
1/Q$
and therefore from dimensional considerations
\begin{equation}
<\,\phi (\overrightarrow{\rho _{1}})...\phi (\overrightarrow{\rho
_{n}^{\prime }})\,>\sim \sum_{n}\int_{-\infty }^{\infty }d\nu \sum_{\mu
_{3}}\sum_{\mu _{4}}...\sum_{\mu _{n}}\,C_{m,\widetilde{m};\mu _{3},...}\,
\frac{\,\Psi _{m,\widetilde{m};\mu _{3},...}(\overrightarrow{\rho }
_{1},...;\,\overrightarrow{\rho }_{0})\,Q^{m}Q^{\widetilde{m}}}{\omega
- \omega (m,\widetilde{m};\mu _{3},...,\mu _{n})}\,.
\end{equation}
It means, that in this limit the contour of integration over $\nu $ should
be shifted to a lower half of the complex plane up to the first pole of
$\left[ \omega -\omega (m,\widetilde{m};\mu _{3},\mu _{2},...,\mu
_{n})\right]^{-1}$.

For small coupling constants $\alpha _{s}$ this singularity
for $n$ reggeized gluons is situated near the pole of
$\omega (m,\widetilde{m};\mu _{3},...,\mu_{n})$.
The position of the leading pole is
\begin{equation}
\frac{m+\widetilde{m}}{2}=\frac{n}{2}-\gamma ^{(n)}\quad , \quad
\gamma^{(n)}=c^{(n)} \; \frac{\alpha _{s}\,N_{c}}{\omega
}\,+O\left(\left[\frac{\alpha_{s}\,N_{c}}{\omega }\right]^{2}\right) \,.
\end{equation}
This relation is in agreement with the above estimate of the anomalous
dimension for the diagrams with the $t$-channel exchange of $p$ 
BFKL Pomerons because here $n=2p$ and $c^{(2p)}=p^{2}$. Moreover, it can
be obtained from the solution of the equation for matrix elements of
quasi-partonic operators written with a double-logarithmic accuracy in ref.
\cite{shuv}. Let us derive a similar equation starting from the
Schr\"{o}dinger equation for the composite states
of reggeized gluons in the multi-colour QCD.

In the case of two reggeized gluons in the momentum space we obtain for
$\left| p_{1}^{\prime }\right| \simeq \left| p_{2}^{\prime }\right| \gg
\left| p_{1}\right| ,\,\left| p_{2}\right| $ the equation
\[
\omega \Psi (\overrightarrow{p}_{1},\overrightarrow{p}_{2})=-\frac{\alpha
_{s}N_{c}\,(\overrightarrow{p}_{1},\overrightarrow{p}_{2})}{\pi \left|
p_{1}\right| ^{2}\left| p_{2}\right| ^{2}}\,\int_{\max (\left| p_{1}\right|
^{2},\,\left| p_{2}\right| ^{2})}^{\infty }d\left| p_{1}^{\prime }\right|
^{2}\int_{0}^{2\pi }\frac{d\varphi _{1^{\prime }}}{2\pi }\,\Psi (
\overrightarrow{p}_{1}^{\prime },-\overrightarrow{p}_{1}^{\prime })\,.
\]
Note, that the Bethe-Salpeter amplitude $\Psi (\overrightarrow{p}_{1},
\overrightarrow{p}_{2})$ contains the gluon propagators $\left| p_{1}\right|
^{-2},\left| p_{2}\right| ^{-2}$. \ The solution of this equation is

\[
\Psi (\overrightarrow{p}_{1},\overrightarrow{p}_{2})\sim \frac{(
\overrightarrow{p}_{1},\overrightarrow{p}_{2})}{\left| p_{1}\right|
^{2}\left| p_{2}\right| ^{2}}\,\left( \frac{\left| p_{1}\right| }{\left|
Q\right| }\right) ^{-\gamma }\,,\,\,\gamma =\frac{\alpha _{s}N_{c}}{\pi
\,\omega }\,.
\]

In the case of $n$ reggeized gluons in the multi-colour QCD apart from the
product of the propagators $\prod_{r=1}^{n}\left| p_{r}\right| ^{-2}$ 
due to the gauge invariance one
can extract from the amplitude also
the momenta $p_{r}^{\mu _{r}}$ for each gluon

\begin{equation}
\Psi (\overrightarrow{p}_{1},\overrightarrow{p}_{2},...,\overrightarrow{p}
_{n})\sim \prod_{r=1}^{n}\frac{p_{r}^{\mu _{r}}}{\left| p_{r}\right| ^{2}}
\,\,f_{\mu _{1},\mu _{2},...,\mu _{n}}\,\,.
\end{equation}
The factor $\prod_{r=1}^{n}p_{r}^{\mu _{r}}/\left| p_{r}\right| ^{2}$ leads
after the Fourier transformation to a singularity in
$\frac{m+\widetilde{m}}{2}$ near $n/2$. The function $f_{\mu _{1},\mu
_{2},...,\mu _{n}}$ is a tensor depending on $\xi _{r}=\ln \,p_{i}^{2}$. 
The
Schr\"{o}dinger equation for this function takes the form

\begin{equation} \omega \,f_{\mu _{1},\mu _{2},...,\mu
_{n}}= \frac{\alpha _{s}N_{c}\,}{4\pi } \sum_{r=1}^{n}\delta _{\mu
_{r}\mu _{r+1}}\delta _{\mu _{r}^{\prime }\mu _{r+1}^{\prime
}}\int_{\xi _{r}}^{\infty }d\xi _{r}^{\prime }\int_{\xi
_{r+1}}^{\infty }d\xi _{r+1}^{\prime }\,\delta (\xi _{r}^{\prime }-\xi
_{r+1}^{\prime })\,\,f_{\mu _{1},\mu _{2},...\mu _{r}^{\prime },\mu
_{r+1}^{\prime }...,\mu _{n}}\,.
\end{equation}

The above equation gives a possibility to calculate
the anomalous dimension $\gamma =\gamma
(\omega )=c_{n}\,\alpha _{s}/\omega $. A similar equation is discussed in
ref.\cite{shuv}. In particular, for the Odderon ($n=3$) it turns out that
$c_{3}=0$ according to an unpublished result of M. Ryskin and A. Shuvaev.
We confirm this result below [see eq.(\ref{mtend1})]
by solving the Baxter equation and finding a
pole singularity near of $\frac{m+\widetilde{m}}{2}=2$ (instead of $3/2$
as it could be expected from above considerations). For
$n=4$ in accordance with the general formula
$\frac{m+\widetilde{m}}{2}=\frac{n}{2}-\gamma ^{(n)}$
we find a pole singularity near $\frac{m+\widetilde{m}}{2}=2$ as
shown in eq.(\ref{quarm2}). 
Moreover, similar to the case of the BFKL
Pomeron the anomalous dimensions $\gamma _{3}$ and $\gamma _{4}$ are
calculated for arbitrary $\alpha /\omega $, which is important
for finding multi-reggeon contributions to the deep-inelastic
processes at small Bjorken's variable $x$. We plot $ \pi \, \omega /[
\alpha \, N_c ] $ as a function  of $ m = \widetilde{m} $ for the Odderon 
in fig.\ref{autm} and for the four-reggeon state in fig. \ref{4regmu}.
It is important, that for $m=\frac{1 \pm k}{2}$ the curves allow
to calculate the intercepts of some states with conformal spin $k$. 

\setlength{\unitlength}{.6 ex}

\section{Solutions of the Baxter equation for the Odderon}

We construct explicit solutions of the Baxter equation for the
Odderon ($n=3$) following the general method presented in sec. 3.

The Baxter equation for the Odderon takes the real form (\ref{Breal})
\bea\label{eqBa3}
&&B_3\left(x;\,m,\mu\right) \equiv \left[ 2 x^3 +
m(m-1)x  +\mu \right] Q\left(x ;\,m,\mu\right)\cr \cr
&&- (x +1)^3 \; Q\left(x +1 ;\, m, \mu\right) -(x-1)^3 \;
Q\left( x-1 ;\,m,\mu\right)=0  \; ,
\eea
where
\be
q_3 = i \mu \,,\,\, Im \, (\mu )=0 \,
\ee
and $\mu$ is assumed to be real, which is compatible with the
single-valuedness condition for the
Odderon wave function in coordinate space \cite{JW}.

The auxiliary functions $ f_r $ (\ref{efe}) for the Odderon take the form
\bea\label{mlodd}
&&f_2\left( x ;\,m,\,\mu\right)=\sum_{l=0}^{\infty} \left[
{a_l(m,\,\mu) \over (x -l)^2}+\,{b_l(m,\,\mu) \over
x -l} \right]\; , \cr \cr
&&f_1\left(x ;\,m,\,\mu\right) = \sum_{l=0}^{\infty} { a_l(m,\mu)
\over x - l } \, .
\eea
Imposing the Baxter equation $ B_3 \equiv 0 $ at the poles $ x = r $
yields the following recurrence relations (cf. \cite{hect})
\bea \label{relrodd}
&&(r+1)^3 \,a _{r+1}(m,\,\mu)=
\left[ 2\,r^3+m(m-1)\,r-\mu\right]\,a _r(m,\,\mu)-(r-1)^3\,
a _{r-1}(m,\,\mu)\; ,     \cr\cr
&&
(r+1)^3 \,b _{r+1}(m,\,\mu)= \left[ 2\,r^3+m(m-1)\,r-\mu \right]\,
b _r(m,\,\mu)-(r-1)^3\,b _{r-1}(m,\,\mu) \cr\cr
&&
+\left[ 6\,r^2+m(m-1) \right]\,a _r(m,\,\mu)-
3\,(r+1)^2\,a _{r+1}(m,\,\mu)-3\,(r-1)^2\,a _{r-1}(m,\,\mu)\; .
\eea
We choose,
\be
a _0(m,\,\mu)=1 \quad , \quad b_0 = 0 \; ,
\ee
and {\bf all } the coefficients $ a _r(m,\,\mu) $ and $ b _r(m,\,\mu)
$ become uniquely determined by eqs.(\ref{relrodd}). In particular,
\bea
&&a _1(m,\,\mu)=-\mu \quad , \quad 8 \,a _2(m,\,\mu)=-\mu
\left[ 2+m(m-1)-\mu\right] \; , \cr\cr
&& b _1(m,\,\mu)=m(m-1)+3 \mu\; , \\ \cr
&& 8 \,b _2(m,\,\mu)= \left[ 2+m(m-1)-\mu \right]\,b _1(m,\,\mu)
- \mu \left[ 6+m(m-1) \right]-12\,a _2(m,\,\mu) \; . \nonumber
\eea
The coefficients 
of the leading pole singularities $a_l(m,\,\mu)  $ 
of $f_1$ and $f_2$ obey
identical recurrence relations as a consequence of the Baxter equation.

Thanks to eq.(\ref{relrodd}), $B_3\left( x ;\,m,\mu\right)$  
for $Q=f_{1,2}$ is
an entire function of $  x $. We find from eq.(\ref{eqBa3}) that it has
generically the form
$$
B_3\left( x ;\,m,\mu\right)= \left[ 6 - m(m-1)\right]
\lim_{x \to \infty} \left[ x \, Q\left( x;\,m,\,\mu\right)\right] \; .
$$
The Baxter equation is fulfilled provided the limit in the r. h. s.
is zero.

\bigskip

Notice that a solution of the Baxter equation multiplied by a periodic
function of $ x $ with its period equal to $ 1 $ is again a solution of 
the Baxter equation. An example of  such a function is $ \pi \cot{\pi x}
$ which is a constant at infinity. Furthermore,
if $ Q\left( x ;\,m,\mu\right) $ is a solution
of eq.(\ref{eqBa3}), then $ Q\left( -x ;\,m,-\mu\right) $ is
also a solution of eq.(\ref{eqBa3}).

\bigskip

We can now form linear combinations  of the auxiliary functions $
f_1\left(x ;\,m,\,\mu\right) $ and $ f_2\left(x ;\,m,\,\mu\right) $
in order to obtain true solutions of the Baxter equation \cite{hect}:
\bea \label{solod}
Q^{(2)}\left(x ;\,m,\mu\right) &=& f_2\left(x ;\,m,\,\mu\right)
+ B(m,\mu)  \; f_1\left(x ;\,m,\,\mu\right) \,,\\ \cr
Q^{(1)}\left(x ;\,m,\mu\right) &=&
f_1\left( x ;\,m,\,\mu\right) +C(m,\mu) \, f_1\left( -x
;\,m,\,-\mu\right)\,, \\ \cr
Q^{(0)}\left(x ;\,m,\mu\right) &=&
Q^{(2)}\left(-x ;\,m,-\mu\right)\,. 
 \nonumber
\eea

As noticed above, the Baxter equation $ B_3\left( x
;\,m,\mu\right) = 0 $ is fulfilled at infinity provided
the coefficients of $ x^{-1} $ in $ Q^{(2)}\left( x
;\,m,\mu\right) $ and in $ Q^{(1)}\left( x ;\,m,\mu\right) $ vanish
for large  $ x $,
\bea \label{b0}
&&\sum_{r=1}^{\infty} b _r(m,\,\mu) + B(m,\mu) \;
\sum_{r=0}^{\infty}a_r(m,\mu) =0 \; ,
\cr \cr
&&\sum_{r=0}^{\infty}a_r(m,\mu) -C(m,\mu)
\,\sum_{r=0}^{\infty} a_r(m,-\mu) = 0 \; .
\eea
This gives for the coefficients the result,
\be
B(m,\mu) = -{\sum_{r=1}^{\infty} b _r(m,\,\mu) \over
\sum_{r=0}^{\infty}a_r(m,\mu)} \quad , \quad
C(m,\mu) ={\sum_{r=0}^{\infty}a_r(m, \mu)\over
\sum_{r=0}^{\infty}a_r(m,-\mu) } \; .
\ee
Therefore, the solutions $Q^{(2)}\left( x ;\,m,\mu\right)$ and
$Q^{(1)}\left( x ;\,m,\mu\right)$ are completely determined.

The functions $Q^{(2)}\left( x ;\,m,\mu\right),\;Q^{(1)}\left( x
;\,m,\mu\right)$ and $ Q^{(0)}\left( x
;\,m,\mu\right) = Q^{(2)}\left( -x
;\,m,-\mu\right) $ 
are related by the linear equation \cite{hect}
\be
\left[ \pi \cot{\pi x} + \delta(m,\mu)
\right] Q^{(1)}\left( x ;\,m,\mu\right) =
Q^{(2)}\left( x ;\,m,\mu\right) -C(m,\mu) \; Q^{(2)}\left(
- x ;\,m,-\mu\right)\,
\ee
where the coefficients in front of $Q^{(2)}\left( x ;\,m,\mu\right)$
and
$Q^{(2)}\left(
- x ;\,m,-\mu\right)$ are calculated from the leading asymptotics at 
$x \rightarrow 1$ and $x \rightarrow -1$, respectively.
This is a special case of eq.(\ref{ecrecu}) for $ n = 3$ and  $ r = 1 $.

By finding the residues at single poles at $x=1$ in the both 
sides of the above equation  one can calculate $\delta(m,\mu)$ 
$$
\delta(m,\mu) = B(m,\mu ) -\frac{C(m,\mu)}{\mu }
\sum_{r=0}^{\infty}\frac{a_r(m,-\mu)}{r+1} -\frac{1}{\mu}
\sum_{r=2}^{\infty }\frac{a_r(m,\mu)}{r-1}-3-
\frac{m(m-1)-1}{\mu} \,.
$$
The total energy for the Odderon described by the function
$ Q_{m,\,\widetilde{m},\,\mu}$ being a bilinear combination
of holomorphic and anti-holomorphic Baxter functions $Q^{(0)}, Q^{(1)}$ 
and $Q^{(2)}$ is  given in ref.\cite{hect}
\begin{equation}
E=i\lim_{\lambda ,\lambda ^{\ast }\rightarrow i}\frac{\partial }{\partial
\lambda }\frac{\partial }{\partial \lambda ^{\ast }}\ln \left[ (\lambda
- i)^{2}(\lambda ^{\ast }-i)^{2}\left| \lambda \right| ^{6}\,
 Q_{m,\,\widetilde{m},\,\mu}
\left( \overrightarrow{\lambda}\right)\right] \,,
\end{equation}
Thus, the energy is expressed in terms of the ratio of residues for
the single and double poles of the Baxter functions at $ x = 1 $. 

The holomorphic energies for the Baxter functions $ Q^{(2)}(x 
,m,\mu)$ and $\pi \cot (\pi x) \, Q^{(1)}(x ,m,\mu)$ should be the same, 
which leads to the quantization of $\mu$ in accordance with 
ref.\cite{hect}
\be \label{autodd}
\delta(m,\mu)=0
\ee
This equation fixes the possible values of $ \mu $ for given $ m$.

Because in the bilinear combination 
$Q_{m,\,\widetilde{m},\,\mu}
\left( \overrightarrow{\lambda}\right)$ there is a 
term proportional to the product of $Q^{(2)}\left(
x ;\,m,\mu\right)$ and $Q^{(2)}\left(
-x^* ;\,\widetilde{m},-\mu \right)$, we obtain for the total
energy 
\bea
E (m,\widetilde{m},\mu) =  B(m,\mu)+B(\widetilde{m},-\mu) \,.
\eea

We found for $m= \widetilde{m}=1/2$ the first roots numerically from the
above
equations (cf. \cite{hect})
\be \label{muodd}
\mu _1 = 0.205257506 \ldots \quad , \quad \mu _2= 2.3439211 \ldots
\quad , \quad\mu _3=8.32635\ldots
\quad , \quad\mu _4=20.080497\ldots 
\ee
with the corresponding energies
\be \label{enodd12}
E_1=0.49434 \ldots \quad , \quad
E_2=5.16930 \ldots \quad , \quad
E_3 = 7.70234 \ldots \quad , \quad
E_4 = 9.46283  \ldots \,.
\ee

\begin{figure}[htbp]
\rotatebox{-90}{\epsfig{file=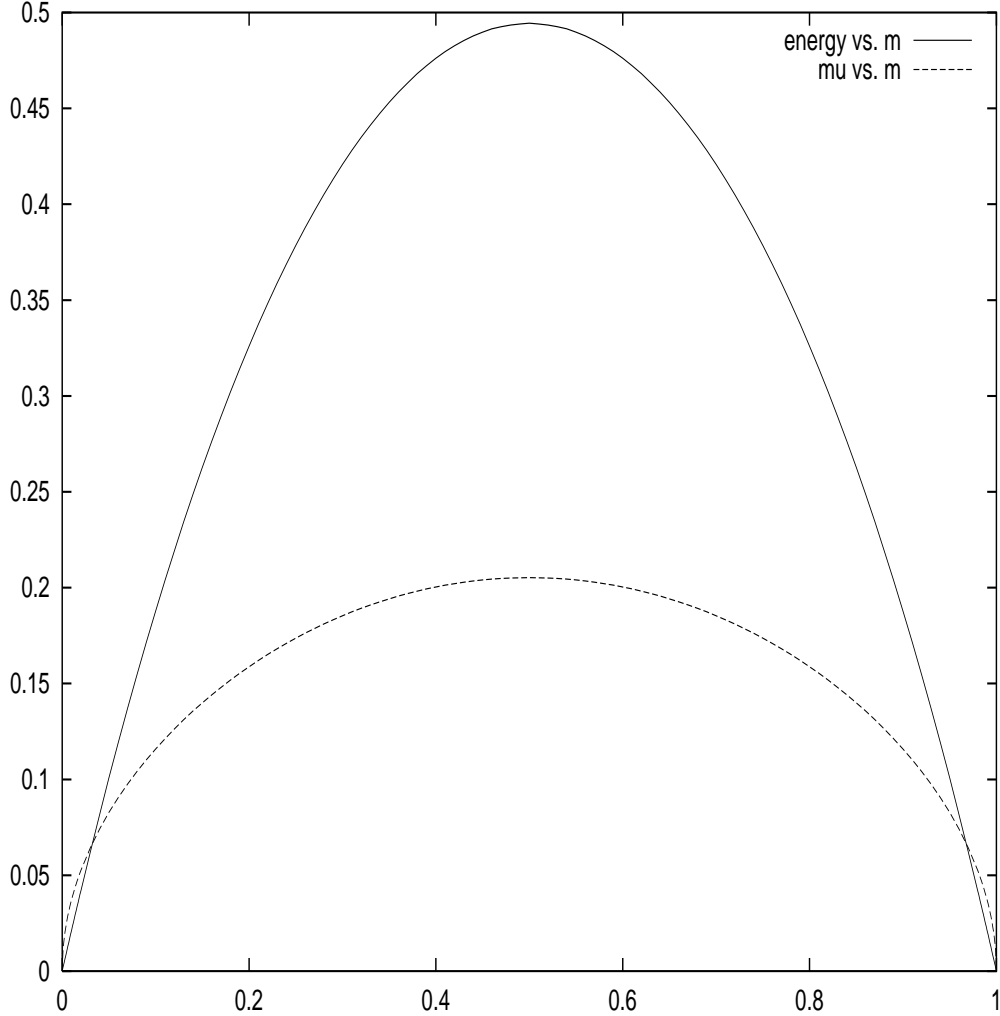,width=14cm,height=14cm}}
\caption{The energy and $\mu $ as functions of $m$ for the odderon
eigenvalue $ E_1, \; \mu_1 $ in the interval $0 < m < 1 $. The picture is
symmetric under $ m \Leftrightarrow 1-m$.}
\label{autm}
\end{figure}

We have followed the eigenvalues $ E_1, \; \mu_1 $ as  functions of $m$ 
for $
0 < m < \frac12 $. The result is plotted in fig. \ref{autm}. Notice
that only $ m = 0, \; 1  $ and $ \frac12 $ are physical values. For other
$m$ the curve describes the behaviour of the anomalous dimension for the
corresponding high-twist operator (see sect. 5).

The energy {\bf vanishes} at $m=0$. This
could be inferred from the fact that $E(m=0,\mu\equiv 0)=0$ in the
expression for  $E(m,\mu\equiv 0)$ given by eq.(74) of ref.\cite{hect}.
$$
E(m,\mu\equiv 0) = \frac{\pi }{\sin (\pi
m)}+\psi (m)+\psi (1-m)-2 \psi (1) \,.
$$
It should be noticed that $E(m,\mu\equiv 0)$ describes an
eigenvalue providing that the function $Q^{(1)}$ does not enter in the
bilinear combination of the total wave function 
$Q_{m,\,\widetilde{m}, \mu}$ because in this case $\delta (m, \mu)$ does 
not vanish.

For the first eigenvalue  we obtain numerically,
\be \label{mchico}
E_1(m) \buildrel{ m \to 0}\over= 2.152\ldots \; m -2.754\ldots \; m^2 +
{\cal O}(m^3) \, , \,\,
\quad \mu_1(m) \buildrel{ m \to 0}\over= 0.375 \ldots \sqrt{m}
- 0.0228 \; m + {\cal O}(m^{\frac32})
\ee

Notice that all quantities are functions of $m$ through the
combination $m(1-m)$. Therefore, they are invariant under the exchange
$ m \Leftrightarrow 1-m $. Thus, eqs.(\ref{mchico}) yield also the
behaviour of $ E_1 $  and $ \mu_1 $ near $ m = 1 $

\bea \label{muno}
E_1(m) &\buildrel{ m \to 1}\over=& 2.152\ldots \; (1-m)  -2.754\ldots
\; (1-m)^2 + {\cal O}\left[ (m-1)^3\right] \; , \cr \cr
\mu_1(m) &\buildrel{ m \to 1}\over= &
0.375\ldots \sqrt{1-m} -0.0228 \; (1-m) + {\cal
O}\left[ (1-m)^{\frac32}\right] \; .
\eea

The state with $m=1$ and $ \widetilde{m} = 0 $ (or viceversa) is
therefore the {\bf ground state} of the Odderon. It has a vanishing energy
and is situated below eigenstates (\ref{enodd12}) with 
$m=\widetilde{m}=1/2$.

\bigskip

One can invert eq.(\ref{muno}) to obtain the anomalous dimension $
\gamma $ [see sec. 5] of the operator with $m=\frac{3}{2}-\gamma $ for
$\omega \rightarrow 0$. The anomalous dimension is not small at arbitrary
$\alpha /\omega$. On the other hand, we can not interpret the state
with $m=3/2$ as a physical state with $n=2, \nu =0$ and a negative energy
(see fig. 3), because $\mu$ is pure imaginary
for this state according to fig. 4. 
Note, that next-to-leading corrections to the BFKL pair kernel  
(cf. \cite{FL}) can give 
contributions to the eigenvalue $\omega$ of the order of $\alpha 
^2/(m-\frac{3}{2})$, which will lead to the usual perturbative expansion
of the anomalous dimension $\gamma \sim \alpha ^2/ \omega$ for the
corresponding  operator.
\bigskip

Furthermore, we consider the states with conformal spin $n=1$
where
$$
m = 1+i \, \nu \quad , \quad\widetilde{m} = i \,  \nu
$$
For small $ \nu $ we compute the energy of such state from
eqs.(\ref{mchico}) and (\ref{muno}) with the result
$$
E = E_1(m) + E_1(\widetilde{m}) = 5.51\ldots\nu^2 + {\cal O} \left(
\nu^4  \right) \; .
$$
The zero energy state that we find for $ \nu = 0 $ is the one found in
ref.\cite{new} by a different approach.

\begin{figure}[htbp]
\rotatebox{-90}{\epsfig{file=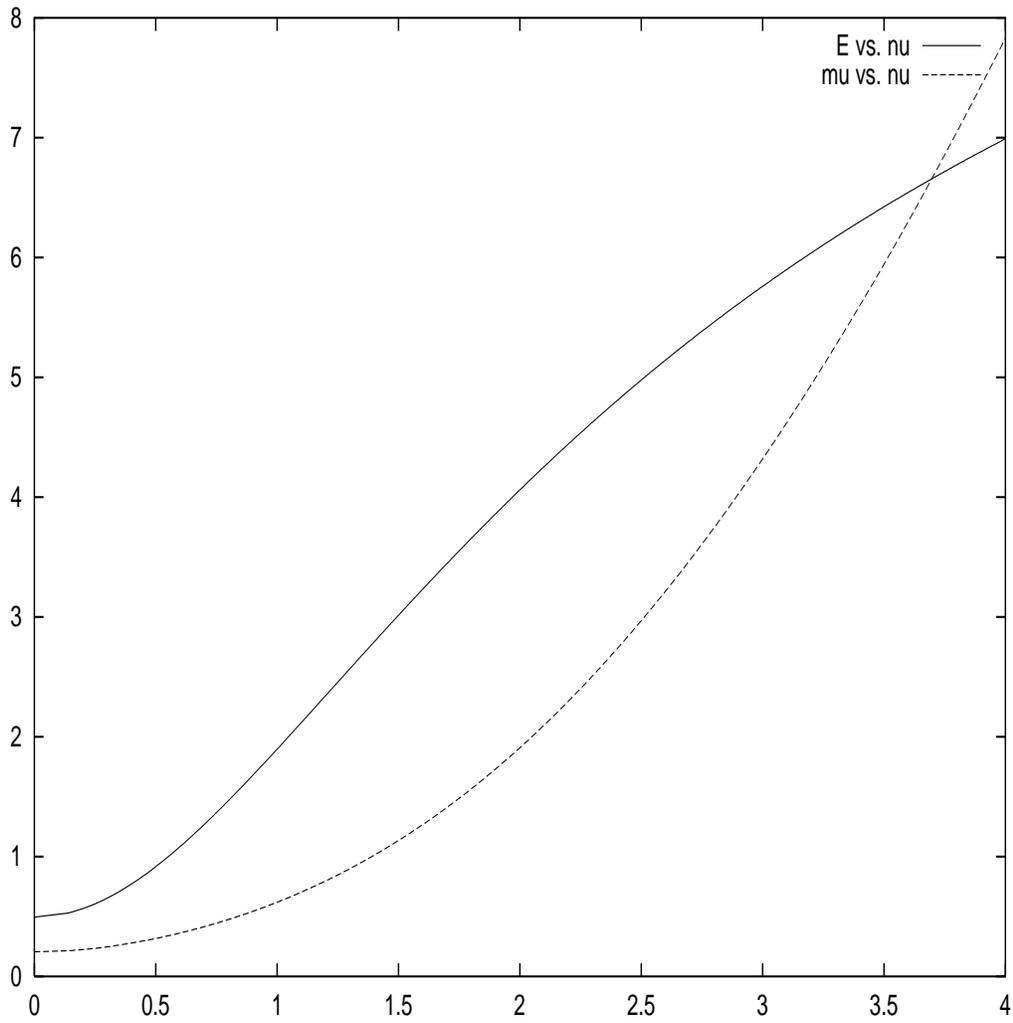,width=14cm,height=14cm}}
\caption{The energy and $ \mu $ as functions of real $\nu$ for the
odderon eigenvalue $ E_1, \; \mu_1 $ setting $m = \frac12 + i \, \nu$.}
\label{autnu}
\end{figure}

We have also followed the first eigenstate  with $n=0$ as a function of $
\nu $
for
real $ \nu $ and
$$
m = \frac12 + i \, \nu
$$
The result is plotted in fig. \ref{autnu}.

The eigenvalues for small $ \nu $ are
$$
E_1(\nu) = 0.49434\ldots + 1.8179\ldots \; \nu^2 + {\cal O}(\nu^4)
\quad , \quad \mu_1(\nu) = 0.205257\ldots + 0.48579\ldots \; \nu^2 +
{\cal O}(\nu^4)
$$

\bigskip

\begin{figure}[htbp]
\rotatebox{-90}{\epsfig{file=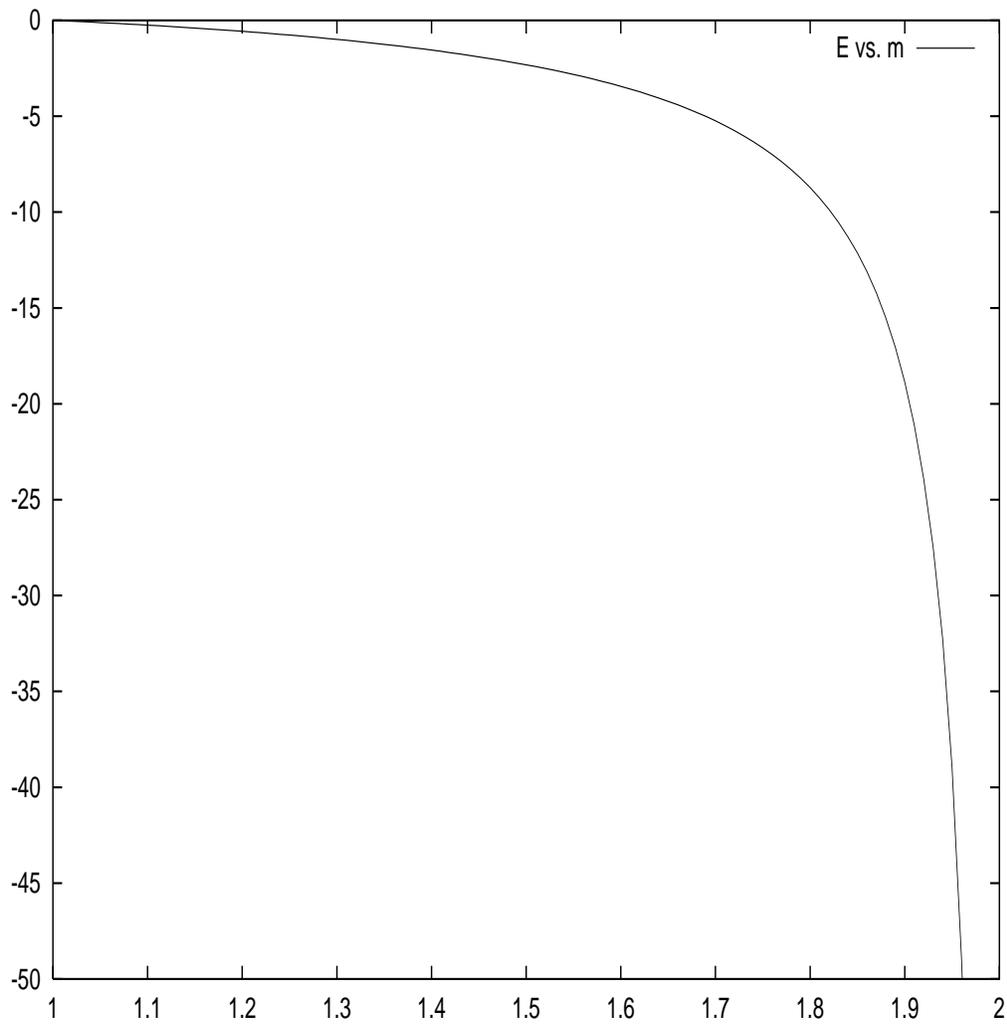,width=14cm,height=14cm}}
\caption{The  odderon energy as a function of $ m $ for the eigenvalue
$1$ in the interval $ 1 \leq m \leq 2$ and imaginary $ \mu $.}
\label{enu}
\end{figure}

\begin{figure}[htbp]
\rotatebox{-90}{\epsfig{file=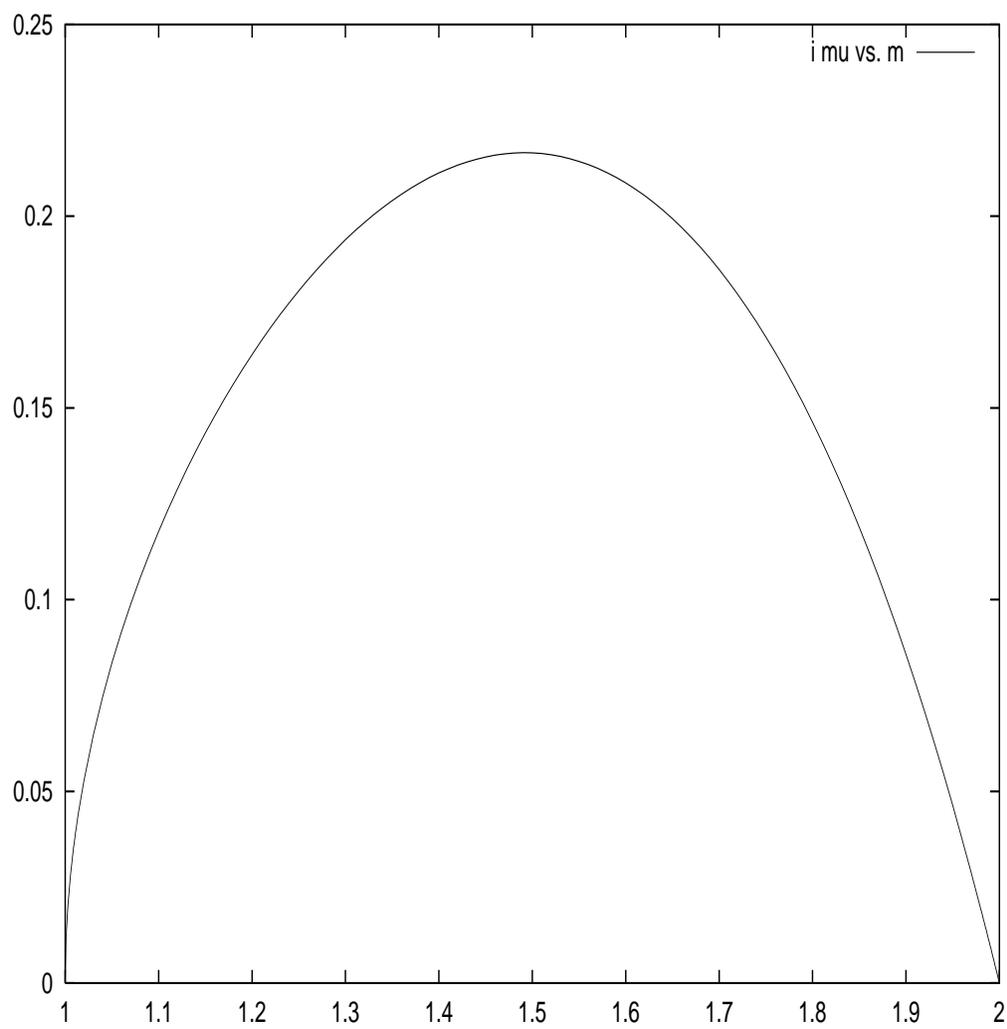,width=14cm,height=14cm}}
\caption{Im$\mu $ as a function of $ m $ for the  odderon eigenvalue $1$
in the interval $ 1 \leq m \leq 2$.[Here Re$\mu=0$].}
\label{imu}
\end{figure}

We continue the first eigenstate with $n=0$ for $ m>1 $ turning $ \mu $ to
purely imaginary. In figs. \ref{enu} and \ref{imu} the energy and
$Im \,\mu $ as functions of $ m $ are plotted for the first eigenvalue  in 
the interval $ 1 \leq m \leq 2$. This describes the dependence of the 
anomalous dimension as a function of $ \alpha_c /{\omega} $ for the 
corresponding operator with $m=2-\gamma$ (see the previous section). 

Near $m=2$ the energy diverges while $ \mu $ tends to
zero. We find its behaviour near $ m=2 $ as follows
\be\label{mtend1}
E_1(m)\buildrel{ m \to 2 }\over= \frac{2}{m-2} + 1 + 2 - m +
{\cal O}\left[ (m-2)^2 \right]
\quad , \quad i\mu \buildrel{ m \to 2 }\over= 2-m - \frac32 \, (m-2)^2
+ {\cal O}\left[ (m-2)^3\right]
\ee
Thanks to the $ m \leftrightarrow 1 - m $ symmetry we have for $
\widetilde{m} \to - 1 $,
$$
E_1 \buildrel{\widetilde{m} \to - 1}\over=-\frac{2}{\widetilde{m}+1} +
\widetilde{m} + 2 +  {\cal O}\left[(\widetilde{m}+1)^2\right] \; .
$$
Now in order to compute the energy of the state with conformal spin $
n = 3 $ we use $ \nu \to 0 $ as a regulator. We have from eqs.(\ref{pesos})
$$
m = 2 + i \nu \quad , \quad \widetilde{m} = -1 + i \nu \; .
$$
Hence,
$$
E = E_1(m) + E_1(\widetilde{m}) \buildrel{ \nu \to 0 }\over= { 2 \over
i \nu} + 1 -  i \nu - { 2 \over i \nu} + 1 +  i \nu + {\cal O}(\nu^2)
= 2 + {\cal O}(\nu^2)
$$

Inverting eq.(\ref{mtend1}) as discussed in sec. 5 yields for the
anomalous dimension for $ \gamma =2-m \rightarrow 0$,
$$
\gamma\buildrel{ \omega \gg \alpha \, N_c }\over= 2 \; { \alpha
\, N_c \over \pi \, \omega}-2 \; \left({ \alpha \, N_c \over \pi \,
\omega}\right)^2- 2 \; \left({ \alpha \, N_c \over \pi \,
\omega}\right)^3+ {\cal O}\left[\left({ \alpha \, N_c \over \pi \,
\omega}\right)^4 \right]\; .
$$

\section{The solutions of the Baxter equation for four reggeons: the
quarteton}

The Baxter equation for the quarteton (four reggeons state)  takes the
form
\bea\label{eqBa4}
&&B_4\left( x ;\,m,\mu,q_4\right) \equiv
\left[ 2 x^4  + m(m-1)x^2  - \mu x + q_4\right]
Q\left( x ;\,m,\mu,q_4\right) \cr \cr
&&-(x +1)^4 \; Q\left(x +1 ;\, m, \mu,q_4\right) - (x -1)^4 \; Q\left( x
- 1 ;\,m,\mu,q_4\right) = 0 \; .
\eea
where $q_3 = i \mu \,,\,\, Im \, (\mu )=0 $.
A new integral of motion $ q_4 $ appears here. $\mu$ and $q_4 $  are
assumed to be real, which is compatible with the single-valuedness of
the wave function in the coordinate space.

Notice that if $ Q\left( x ;\,m,\mu,q_4\right) $ is a solution
of eq.(\ref{eqBa4}) then $ Q\left( -x ;\,m,-\mu,q_4\right) $ is
also a solution of eq.(\ref{eqBa4}).

\bigskip

Following the general method presented in sec. III and \cite{hect}
we seek solutions of the
Baxter equation for the quarteton as a series of poles. We start by
finding
recurrence relations for the residues of the poles. Then, we
impose the validity of the Baxter equation at infinity which
gives further linear constraints on the pole residues.

The auxiliary functions (\ref{efe}) take here the form
\bea\label{elem4}
&&f_3\left( x ;\,m,\,\mu,q_4\right)=\sum_{l=0}^{\infty} \left[
{a _l(m,\,\mu,q_4) \over (x-l)^3}+{b _l(m,\,\mu,q_4) \over
(x-l)^2 } + {c_l(m,\,\mu,q_4) \over x-l }
\right]\, ,\cr \cr
&&f_2\left( x ;\,m,\,\mu,q_4\right)=\sum_{l=0}^{\infty} \left[
{a _l(m,\,\mu,q_4) \over (x-l)^2}+{ b_l(m,\,\mu,q_4) \over
x-l } \right] \, , \cr \cr
&&f_1\left( x ;\,m,\,\mu,q_4\right)=\sum_{l=0}^{\infty}
{a _l(m,\,\mu,q_4) \over x-l} \; .
\eea
Imposing the Baxter equations {\bf at} the poles $ x= l \;
, l=0,1,2,3,\ldots $ yields the recurrence relations:
\bea\label{rr4}
&&(l+1)^4 \,a_{l+1}(m,\,\mu,q_4)=
\left[ 2\,l^4+m(m-1)\,l^2-\mu \, l \; + q_4 \right]\,a
_l(m,\,\mu,q_4)-(l-1)^4\,
a _{l-1}(m,\,\mu,q_4)\; ,     \cr\cr
&&
(l+1)^4 \,b _{l+1}(m,\,\mu,q_4)= \left[ 2\,l^4+m(m-1)\,l^2-\mu\, l+
q_4  \right]\,
b_l(m,\,\mu,q_4)-(l-1)^4\,b_{l-1}(m,\,\mu,q_4) \cr\cr
&&
+\left[ 8\,l^3+2m(m-1)\, l -\mu \right]\,a _l(m,\,\mu,q_4)-
4\,(l+1)^3\,a _{l+1}(m,\,\mu,q_4)-4\,(l-1)^3\,a _{l-1}(m,\,\mu,q_4)\;
,\cr\cr
&&
(l+1)^4 \,c_{l+1}(m,\,\mu,q_4)= \left[ 2\,l^4+m(m-1)\,l^2-\mu\, l+ q_4
\right]\,
c_l(m,\,\mu,q_4)-(l-1)^4\,c_{l-1}(m,\,\mu,q_4) \cr\cr
&&
+\left[ 8\,l^3+2m(m-1)\, l -\mu \right]\, b_l(m,\,\mu,q_4)-
4\,(l+1)^3\,b_{l+1}(m,\,\mu,q_4)-4\,(l-1)^3\,b_{l-1}(m,\,\mu,q_4) \cr\cr
&&
- 6(l+1)^2 \, a_{l+1}(m,\,\mu,q_4) - 6(l-1)^2 \, a_{l-1}(m,\,\mu,q_4) + [12
l^2
+ m(m-1) ]a _l(m,\,\mu,q_4) \; .
\eea
We choose,
\be\label{condin}
a _0(m,\,\mu,q_4)=1 \quad , \quad b_0 = 0 \quad , \quad c_0 = 0\; ,
\ee
and {\bf all } the coefficients $ a _l(m,\,\mu,q_4) , \;
b_l(m,\,\mu,q_4)$ and $ c_l(m,\,\mu,q_4) $ become uniquely determined
by eqs.(\ref{rr4}). In particular,
\bea
&&a_1(m,\,\mu,q_4)= q_4 \quad , \quad
b _1(m,\,\mu,q_4)=-4 q_4 - \mu\; , \cr\cr
&& c_1(m,\,\mu,q_4)= 10 \, q_4 + 4\mu + m(m-1) \; .
\eea

Taking linear combinations of $ f_1\left( \pm x
;\,m,\,\pm\mu,q_4\right) , f_2\left( x ;\,m,\,\mu,q_4\right) $ and
$ f_3\left( x ;\,m,\,\mu,q_4\right) $ as in eqs.(\ref{solnm1}) and
(\ref{solt}) we form three independent solutions $Q^{(r)}$ of the Baxter
equation (\ref{eqBa4}):
\bea\label{solu4}
Q^{(3)}\left( x ;\,m,\,\mu,q_4\right) &\equiv& f_3\left( x
;\,m,\,\mu,q_4\right) + \alpha_1(m,\mu,q_4) \; f_2\left( x
;\,m,\,\mu,q_4\right) + \alpha_2(m,\mu,q_4) \; f_1\left( x
;\,m,\,\mu,q_4\right)\, ,\cr \cr
Q^{(2)}\left( x ;\,m,\,\mu,q_4\right) &\equiv&
f_2\left( x ;\,m,\,\mu,q_4\right) + \gamma_1(m,\mu,q_4) \;
f_1\left( x ;\,m,\,\mu,q_4\right) +  \cr \cr
&+&\gamma_2(m,\mu,q_4) \; f_1\left( -x ;\,m,\,-\mu,q_4\right) \; .
\eea
The solution $ Q^{(1)}\left( x ;\,m,\,\mu,q_4\right) $ is proportional
to $ Q^{(2)}\left( -x ;\,m,\,-\mu,q_4\right) $.

The recurrence relations for the coefficients of
the poles guarantee that $ B_4\left( x ;\,m,\mu,q_4\right) $ is
an entire function. More precisely, we find from
eqs.(\ref{eqBa4}) and (\ref{elem4}) that $ B_4\left( x
;\,m,\mu,q_4\right) = k_1 + k_2 \, x $, where $k_1$ and $k_2$
are some constants. These constants vanish and the Baxter equation is
fulfilled {\bf provided}  the coefficients of $ x^{-1} $ and  $
x^{-2} $ vanish for large $ x $ in the $Q'$s of eq.(\ref{solu4}).
Because the coefficients of $ x^{-1} $ and $ x^{-2} $ according to
eqs.(\ref{asi}) from Appendix A are nonzero for the auxiliary functions $ 
f_1\left( x
;\,m,\,\mu,q_4\right) , \; f_2\left( x ;\,m,\,\mu,q_4\right) $
and $ f_3\left( x ;\,m,\,\mu,q_4\right) $, these functions {\bf are not}
solutions of the Baxter equation.

The coefficients $ \alpha_1(m,\mu,q_4), \; \alpha_2(m,\mu,q_4),
\;\gamma_1(m,\mu,q_4) $ and $ \gamma_2(m,\mu,q_4) $ are chosen
imposing the Baxter equation at infinity.
We present the linear equations on $ \alpha_1(m,\mu,q_4), \;
\alpha_2(m,\mu,q_4) $ , $\gamma_1(m,\mu,q_4) $ and $ \gamma_2(m,\mu,q_4)
$ and their explicit solutions in  Appendix A.

\bigskip

There is one linear relation (\ref{ecrecu}) among the
Baxter solutions $ Q^{(t)} $ for $ n = 4 $
$$
\left[ \delta^{(2)}(\mu,\,m,q_4) + \pi
\cot{\pi x}  \right] Q^{(2)}\left( x ;\,m,\,\mu,q_4\right)
= Q^{(3)}\left(
x ;\,m,\,\mu,q_4\right) + \alpha^{(2)}(\mu,\,m, q_4) Q^{(2)} \left(
- x ;\,m,\,-\mu,q_4\right) \; .
$$
where $\delta^{(2)}(\mu,\,m,q_4) = \alpha_1(\mu,\,m,q_4) -
\gamma_1(\mu,\,m,q_4) $. A proof of this equation is given
in Appendix B. Another relation is obtained from above one
by the substitution
$x \rightarrow -x, \, \mu \rightarrow -\mu$.

The energy of the four Reggeons state is expressed through the bilinear
combination $Q_{m,\,\widetilde{m},\,\mu ,\, q_4}$ of the Baxter functions
\cite{hect}
\bea\label{ener4}
E=i\lim_{\lambda ,\lambda ^{\ast }\rightarrow i}\frac{\partial }{\partial
\lambda }\frac{\partial }{\partial \lambda ^{\ast }}\ln \left[ (\lambda
- i)^3(\lambda ^{\ast }-i)^3\left| \lambda \right| ^{8}\,
 Q_{m,\,\widetilde{m},\,\mu ,\, q_4}
\left( \overrightarrow{\lambda}\right)\right] \,,
\eea 
Again, $E$ is related to the behavior of the Baxter
functions $Q^{(r)}$ near $ x = 1 $. That is,
\bea\label{lami}
&&Q^{(3)}\left( x ;\,m,\,\mu,q_4\right)\buildrel{ x \to
1}\over = {q_4 \over (x -1)^3}
\left[ 1 - (x -1) \left(4+ {\mu \over q_4} +
\alpha_1(m,\mu,q_4) \right) + {\cal O}\left([x -1]^2\right)\right],\cr \cr
&&\pi \, \cot[\pi x] \; Q^{(2)}\left( x ;\,m,\,\mu,q_4\right)\buildrel{ x
\to
1}\over= -{q_4 \over (x - 1)^3}
\left[ 1 - (x - 1) \left(4+ {\mu \over q_4} +
\gamma_1(m,\mu,q_4) \right) + {\cal O}\left([x - 1]^2\right)\right].
\nonumber \\
\eea
We obtain for the total energy of the three solutions from 
eqs.(\ref{ener4}) and (\ref{lami}),
\bea\label{e3ye2}
E^{(3)}(m,\widetilde{m}, \mu,q_4) &=&-
\alpha_1(m,\mu,q_4)-\alpha_1(\widetilde{m},-\mu,q_4) \; , \cr \cr
E^{(2)}(m,\widetilde{m},\mu,q_4) &=&-
\gamma_1(m,\mu,q_4)-\gamma_1(\widetilde{m},-\mu,q_4) \; .
\eea
The eigenvalue conditions
\be\label{autov}
\delta ^{(2)}(\mu , m, q_4) \sim \alpha_1(\mu,\,m,q_4) - 
\gamma_1(\mu,\,m,q_4) = 0 \quad , \quad
\delta ^{(2)}(-\mu,\,\widetilde{m}, q_4) = 0
\ee
guarantee that the holomorphic energy is the same for these three independent
Baxter solutions,
$$
\epsilon ^{(3)}(m, \mu,q_4) =
\epsilon ^{(2)}(m,\mu,q_4)=\epsilon ^{(1)}(m,\mu,q_4)\,.
$$
Eqs.(\ref{autov})  fix possible values of $ \mu $ and $ q_4 $ for
given $ m$.

We find from the above
equations for $m= \widetilde{m}=1/2$  the first roots numerically as,
\bea \label{lista}
&&\mu = 0 \quad ,  \quad  q_4= 0.1535892\ldots \quad ,
\quad  E= -1.34832\ldots \quad , \cr \cr
&&\mu = 0.73833\ldots \quad ,  \quad  q_4= -0.3703\ldots \quad ,
\quad  E= 2.34105\ldots \quad , \cr \cr
&&\mu = 0 \ldots \quad ,  \quad  q_4= -0.292782\ldots \quad ,
\quad  E= 2.756624\ldots \quad , \cr \cr
&&\mu = 1.4100 \ldots \quad ,  \quad  q_4= 0.73852\ldots \quad ,
\quad  E= 3.3581\ldots \quad , \cr \cr
&&\mu = 0 \ldots \quad ,  \quad  q_4= 1.79992\ldots \quad ,
\quad  E= 5.67117\ldots \quad , \cr \cr
&&\mu = 0 \ldots \quad ,  \quad q_4=-2.1857905 \ldots \quad ,
\quad  E=6.2819490\ldots \quad .
\eea
The first eigenstate was reported in ref.\cite{num} where it is
incorrectly identified as the ground state for the quarteton [see nota added].

In fig.\ref{curva} the eigenvalue equations (\ref{autov}) 
are plotted in
the $ \mu, \; q_4 $-plane for $ m = \widetilde{m} = \frac12 $. The
curves intersect at the eigenstates  (\ref{lista}).

We find for the first eigenvalues with $ m = 0, \;   \widetilde{m} = 1
$ corresponding to $ n = -1 $ in eq.(\ref{pesos}).[ $ n = 1 $ gives the same
state with $ m \leftrightarrow \widetilde{m} $].
\bea \label{lista0}
&&\mu = 0 \quad ,  \quad  q_4= 0.12167\ldots \quad ,
\quad  E= -2.0799\ldots \quad , \cr \cr
&&\mu = 0.51214\ldots \quad ,  \quad  q_4= -0.33288\ldots \quad ,
\quad  E= 2.2007\ldots \quad , \cr \cr
&&\mu = 0 \quad ,  \quad  q_4=-0.2905426 \ldots \quad ,
\quad  E= 2.441210\ldots \quad .
\eea
Therefore, the {\bf ground state} of the quarteton 
with $|n|=1$ corresponds to $ m
= 0, \;   \widetilde{m} = 1 $. Its energy, $ E = -2.0799\ldots $ is
{\bf below } the lowest energy $  E= -1.34832\ldots $ of the 
states with $ m = \widetilde{m} = \frac12 $. 

We plot in fig. \ref{curva0} the eigenvalue equations (\ref{autov}) in
the $ \mu, \; q_4 $-plane for $ m = 0, \;   \widetilde{m} = 1 $. The
curves intersect at the eigenstates  (\ref{lista0}). It should be
noticed that the curves 
in figs. \ref{curva} and \ref{curva0} turn to be
qualitatively similar. The eigenstates with  $ m = 0, \;   \widetilde{m} = 1 $ 
 follow from those with $ m = \widetilde{m} = \frac12 $ by an analytic
continuation in $ m, \; \widetilde{m}$.

\begin{figure}[htbp]
\rotatebox{-90}{\epsfig{file=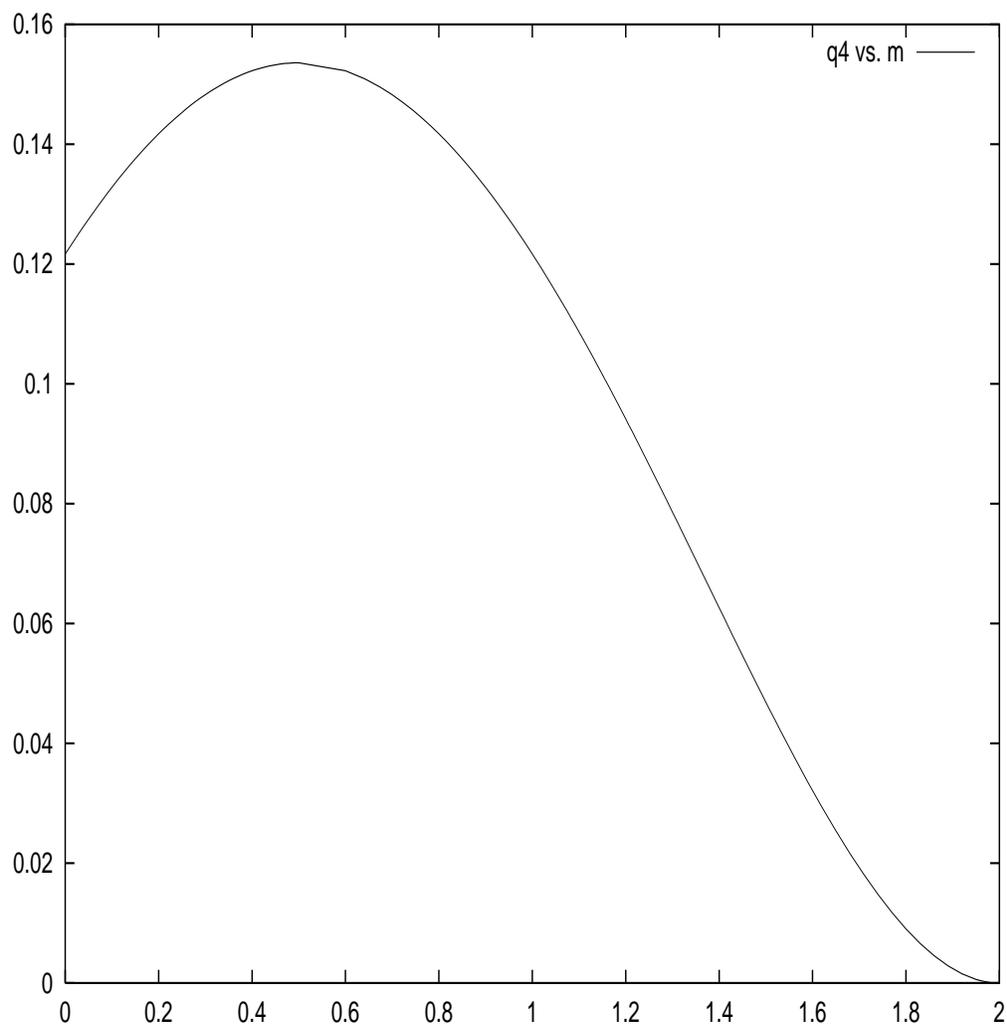,width=14cm,height=14cm}}
\caption{The $ q_4 $ as a function of $m$ for the quarteton eigenvalue
$1$ in the interval $0 < m < 2 $. We have for this eigenvalue $ \mu
= 0 $ for all $m$. }
\label{4regmu}
\end{figure}

\begin{figure}[htbp]
\rotatebox{-90}{\epsfig{file=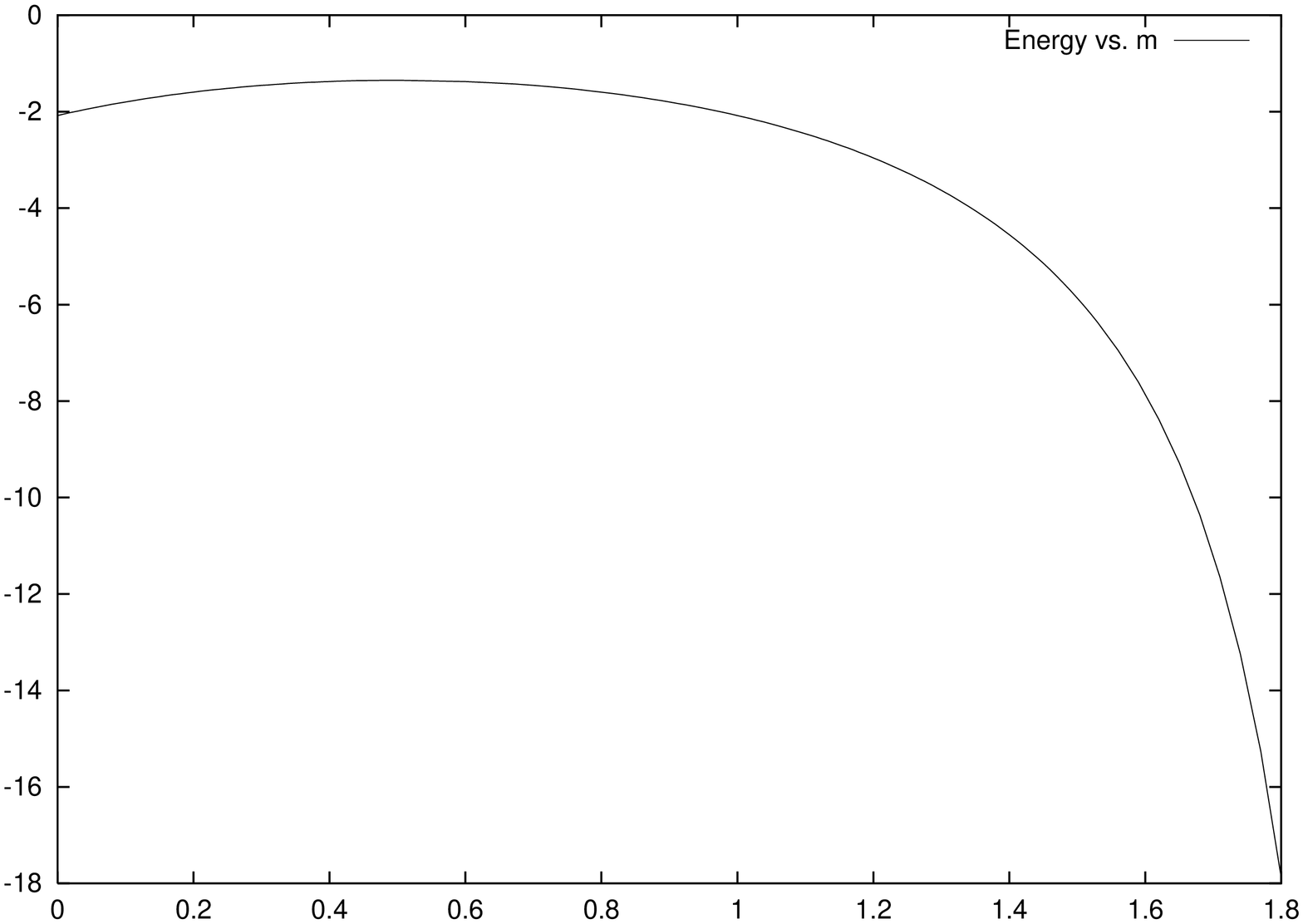,width=14cm,height=14cm}}
\caption{The energy as a function of $m$ for the  quarteton eigenvalue
$1$ in the interval $0 < m < 2 $. We have for this eigenvalue $ \mu
= 0 $ for all $m$. }
\label{4rege}
\end{figure}

\begin{figure}[htbp]
\rotatebox{-90}{\epsfig{file=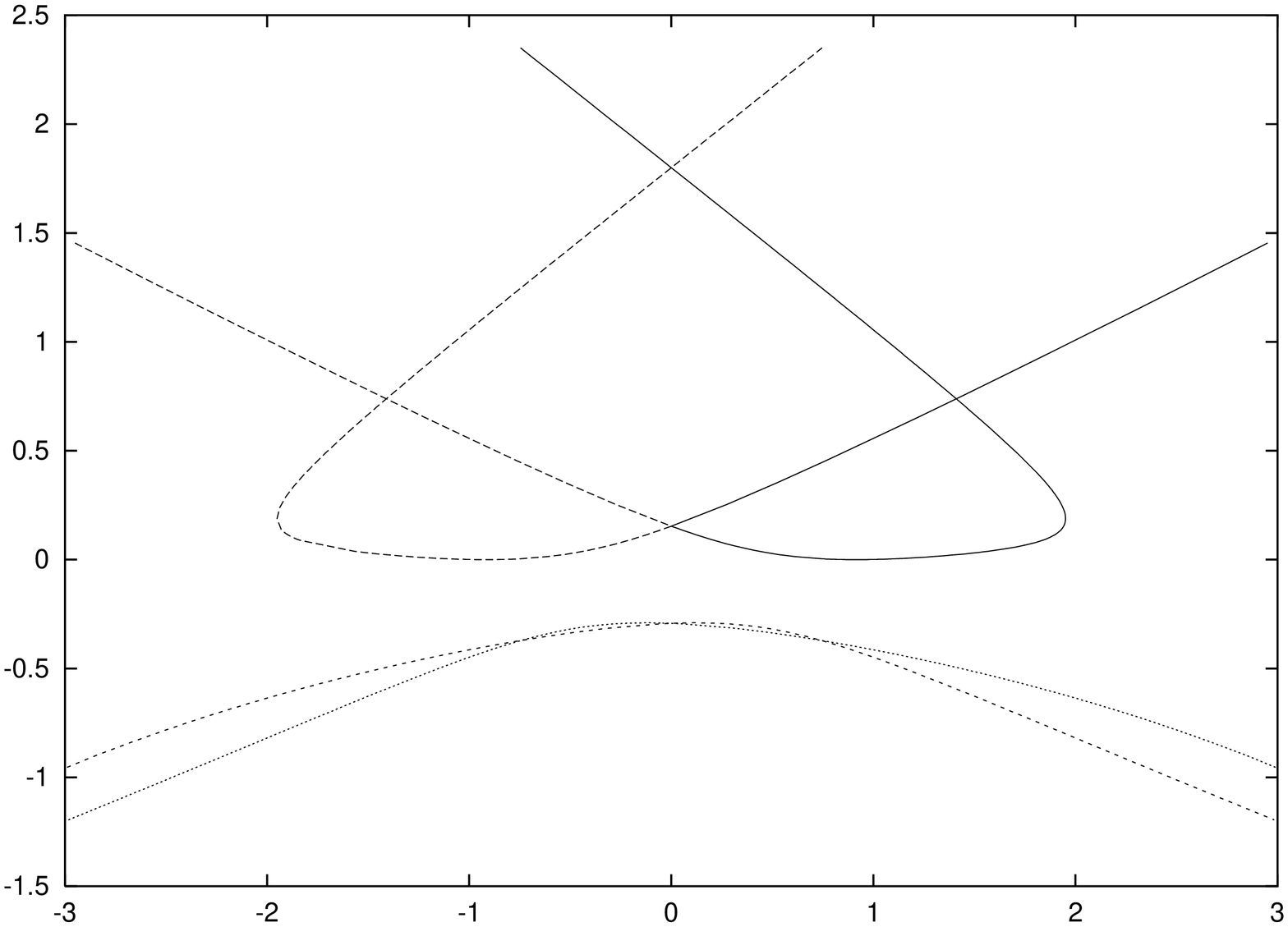,width=14cm,height=14cm}}
\caption{The  quarteton eigenvalue equations $ \alpha_1(\mu,\,m,q_4) -
\gamma_1(\mu,\,m,q_4) = 0 $ and $\alpha_1(-\mu,\,\widetilde{m},q_4) -
\gamma_1(-\mu,\,\widetilde{m},q_4) = 0 $ for $ m = \widetilde{m} =
\frac12 $ in the $ \mu, \; q_4 $-plane }
\label{curva}
\end{figure}

\begin{figure}[htbp]
\rotatebox{-90}{\epsfig{file=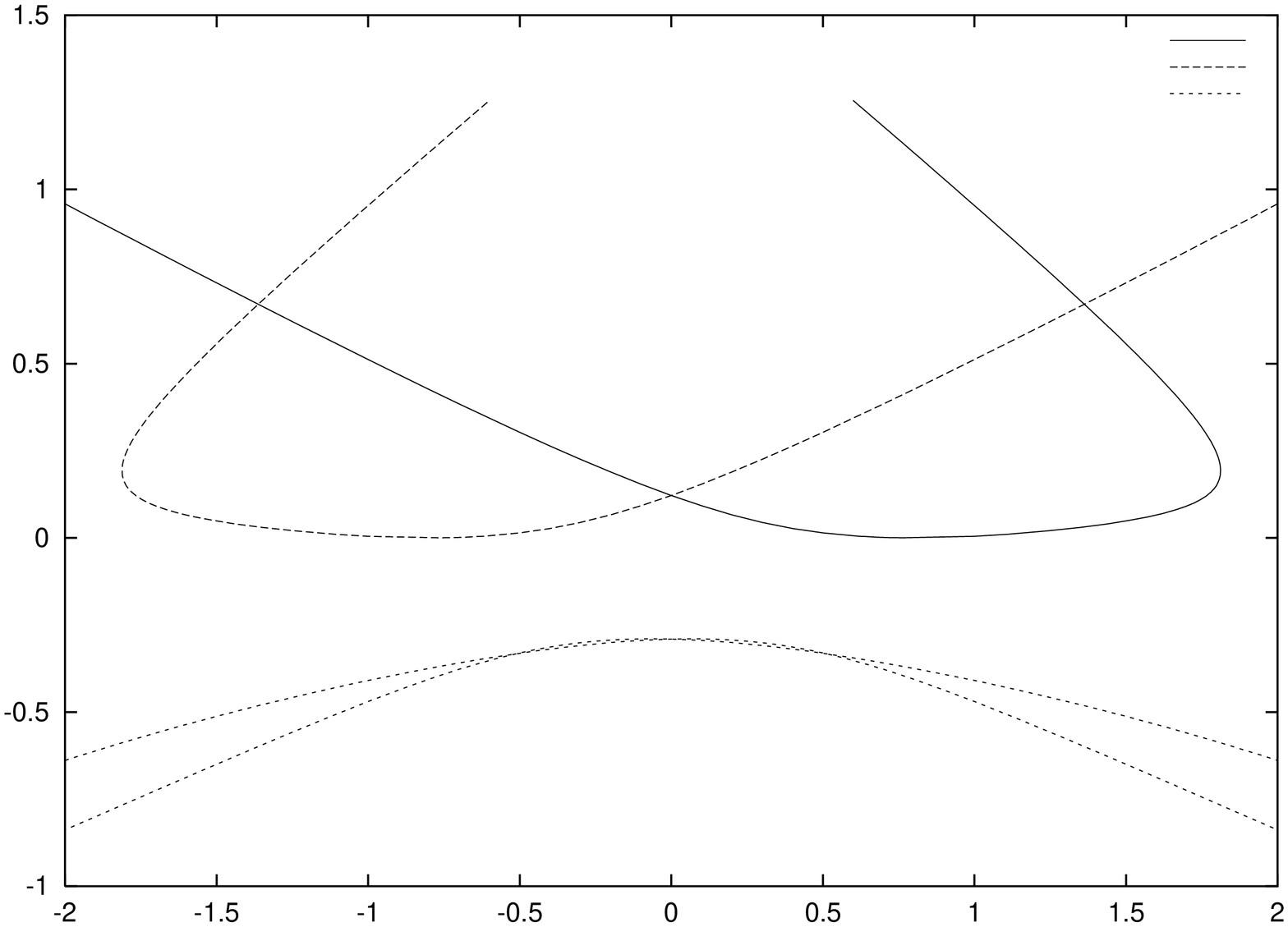,width=14cm,height=14cm}}
\caption{The  quarteton eigenvalue equations $ \alpha_1(\mu,\,m,q_4) -
\gamma_1(\mu,\,m,q_4) = 0 $ and $\alpha_1(-\mu,\,\widetilde{m},q_4) -
\gamma_1(-\mu,\,\widetilde{m},q_4) = 0 $ for $ m = 0, \; \widetilde{m} =
 1  $ in the $ \mu, \; q_4 $-plane }
\label{curva0}
\end{figure}

We have followed the first eigenvalue as a function of $m$ for $
0 < m < \frac12 $. The result is plotted in fig. \ref{4rege}.
Contrary to the Odderon case, the energy eigenvalue does
not vanish for $ m = 0 $. The energy decreases with $m$ for $
0 < m < \frac12 $ and takes the value $ E = -2.0799 \ldots $ at $ m = 0
$ while $ q_4 $ takes there the value $ 0.12167 \ldots $. Further, $ \mu $
vanishes
for this eigenvalue for all $m$ [see fig. \ref{4regmu}]. 

For $ m = 2 $ the lowest energy state of four reggeons goes to minus
infinity and $q_4$ vanishes  [$q_3$ is zero for all m in this
state]. We find the following behaviour, 
\be\label{quarm2}
E_1 \buildrel{ m \to 2}\over= \frac{4}{m-2} + 2 + 2-m + {\cal
O}\left[(m-2)^2 \right]  \quad , \quad
q_4 \buildrel{ m \to 2}\over= \frac14 (m-2)^2 +  {\cal O}\left[(m-2)^3
\right] \; .
\ee
Thanks to the $ m \leftrightarrow 1 - m $ symmetry we have for $
\widetilde{m} \to - 1 $,
$$
E_1 \buildrel{\widetilde{m} \to - 1}\over=-\frac{4}{\widetilde{m}+1} +
3 + \widetilde{m}  +  {\cal O}\left[(\widetilde{m}+1)^2\right] \; .
$$
Now in order to compute the energy of the state with conformal spin $
n = 3 $ we use $ \nu \to 0 $ as a regulator as in the odderon case.
We find in this way,
$$
E = E_1(m) + E_1(\widetilde{m}) \buildrel{ \nu \to 0 }\over= 4 + {\cal
O}(\nu^2) 
$$
Inverting eq.(\ref{quarm2}) as discussed in sec. 5 yields for the 
anomalous dimension for $ \gamma =2-m \rightarrow 0 $,
$$
\gamma\buildrel{ \omega \gg \alpha \, N_c }\over=  4 \; { \alpha
\, N_c \over \pi \, \omega} +8 \; \left({ \alpha \, N_c \over \pi \,
\omega}\right)^2 + {\cal O}\left[ \left({ \alpha \, N_c \over \pi \,
\omega}\right)^4 \right]\; .
$$
 
The state with $m=3/2$ (with $n=2,\, \nu =0$) can be considered
as a physical ground state for the quarteton  because for it the
eigenvalue 
of $q_4$ is real according to fig. 5. It has a large negative energy
comparable with the energy of the BFKL pomeron constructed from two
reggeized gluons. But to prove this conclusion it is needed to construct
a bilinear
combination of the corresponding Baxter functions to verify the
normalizability of the corresponding solution.  

\vspace{1cm}
\begin{tabular}{|l|l|l|l|l|}\hline
$$ & $$ & energy & energy  & energy  \\
conformal spin & conformal weights & eigenvalues  &eigenvalues  & eigenvalues\\
$$ & $$ & pomeron & odderon &quarteton  \\ \hline 
$$ & $$ & $$ & $$ &  \\
$n=0$&$m=\widetilde{m}=1/2$& $-8 \; \ln2=-5.545177$&$0.49434$&$-1.34832$ \\
$$ & $$ & $$ & $$ &  \\
$n=1$ & $m=1, \; \widetilde{m}=0$ & $\hspace{0.6cm}0$ &
$\hspace{0.6cm}0$ & $-2.0799$ \\ 
$$ & $$ & $$ & $$ &  \\
$n=2$ & $m=3/2, \;  \widetilde{m} = -1/2$ &  $8(1-\ln 2)=2.45482$ &
imaginary $\mu$ & $-5.863$ \\ 
$$ & $$ & $$ & $$ &  \\
$n=3$ & $m=2, \;  \widetilde{m} = -1$ &  $\hspace{0.6cm}4$ &
\hspace{0.6cm}  $2$ & \hspace{0.6cm} $4$ \\ 
$$ & $$ & $$ & $$ & \\ \hline 
\end{tabular}

\vspace{0.5cm}

{TABLE 1. Conformal spins, conformal weights and corresponding
lowest energy eigenvalues for the pomeron, odderon and quarteton
states. The odderon state with imaginary $\mu$ is discarded as non-physical.
The pomeron and quarteton states with $n=3$ have the same energy and are
related presumably by the duality symmetry.}

\bigskip

As displayed in Table 1 the intercept for the quarteton ground state
possessing conformal spin $n=2$ is larger than that for the
BFKL Pomeron. This result is not very suprising because four
reggeized gluons clustered into two pomerons have an even larger
intercept. A large intercept for the state with the conformal spin 2 may
lead to such unphysical
results as negative cross sections. But it is known that the
unitarization of scattering amplitudes is not solved within a framework
where the number of reggeons is fixed. 

\section{Acknowledgements}

We thank E. A. Antonov, J. Bartels, A. A. Belavin, A. P. Bukhvostov,
R. Kirschner, E. A. Kuraev, G. Marchesini, G. P. Vacca, W. von
Schlippe and further
participants to the PNPI Winter School for stimulating discussions on the
basic results of this paper. Subsequent discussions with 
V. Fateev, P. Mitter. A. Neveu, F. Smirnov and Al. Zamolodchikov were
especially fruitful. One of us (LNL) thanks LPTHE for the hospitality
during his visit to the University  of Paris VI in March  2002.             

\bigskip

{\bf Nota Added}: After completion of this work we have seen the
preprint \cite{num2} studying similar problems. 

It is stated in refs.\cite{num,num2} that the ground state for
three and four reggeons corresponds to vanishing conformal spin $ m -
\widetilde{m} = 0 $. We show here that the lowest energy eigenvalue
corresponds in both cases to  a non-zero conformal spin $n = 1$ and
$n=2$, respectively. 

It is claimed in ref.\cite{num2} that the eigenvalues $\mu$ 
can be complex. In our opinion the reason for this disagreement 
with our results is, that
the authors of ref.\cite{num2} did not take into account, that in the 
Baxter-Sklyanin representation there are two different expressions
for the energy in terms of $Q_{m}(\lambda), 
Q_{\widetilde{m}}(\lambda ^{*})$ obtained at $\lambda \rightarrow \pm i$. 
These  two energy values coincide if and only if $\psi 
_{m,\widetilde{m}}$ is an eigenfunction of the Hamiltonian. It leads to 
the  constraint eq.(\ref{condmu}) on the spectrum of the integrals of 
motion.  

We have computed the energies for the odderon states displayed in
table 1 of ref.\cite{num2}. States $(0,2)$ and $(6,2)$ coincide with
our first two eigenstates $1$ and $2$ [see eqs.(\ref{muodd}) and
(\ref{enodd12})] within the 
three figures precision of table 1 of ref.\cite{num2}. States $
(2,2), \; (4,2), \; (8,2) $ and $ (10,2) $ possess complex $\mu$ and
complex energy. Moreover, they fulfill the real part of our eigenvalue equation
(\ref{autodd}) up to $ {\cal O}(10^{-3}) $ while the imaginary part of 
$ \delta(m,\mu) $ turns to be of the order one. Our understanding is
that these states with complex $ \mu $ {\bf are not} eigenstates of
the hamiltonian.  

We furthermore computed the energies for the quarteton states displayed in
table 2 of ref.\cite{num2}. States $(2,0), \; (2,2), \; (4,0) $ and $
(3,3) $ coincide with our quarteton eigenvalues [see eq.(\ref{lista})]
1, 3, 5 and 6, respectively. (Again, within the three figures
precision of table 2 of ref.\cite{num2}). States $(3,1)$ and $(4,2)$
possess complex $ q_4 $ and complex energy. The real parts of $E^{(3)}$
and $E^{(2)}$ computed from eqs.(\ref{e3ye2}) coincide up to $ {\cal
O}(10^{-3}) $ while the imaginary parts of $E^{(3)}$ and $E^{(2)}$ differ by
amounts of the order one. Our understanding is again
that these states with complex $  q_4 $ {\bf are not} eigenstates of
the hamiltonian.  

\appendix

\section{Asymptotic constraints on the solutions of the
Baxter equation. The quarteton case.}

We impose here the Baxter equation at infinity on the solutions for four
reggeons. As derived in sec. VI, the coefficients of $ x^{-1} $ and $
x^{-2} $ in $ Q^{(3)}\left( x ;\,m,\,\mu,q_4\right) $ and  $
Q^{(2)}\left( x ;\,m,\,\mu,q_4\right) $ for $ x \to \infty $ must vanish.

For large $ x $ the series (\ref{elem4}) yield the following formal
expansions,
\bea\label{asi}
&&f_3\left( x ;\,m,\,\mu,q_4\right)\buildrel{ |x| \to
\infty}\over= { 1 \over ix } \sum_{r=0}^{\infty}c_r(m,\,\mu,q_4)
+ { 1 \over (ix)^2 } \sum_{r=0}^{\infty}\left[ b_r(m,\,\mu,q_4)
-  r \,c_r(m,\,\mu,q_4) \right] + {\cal O}\left({ 1 \over x^3
}\right)\, ,\cr \cr
&&f_2\left( x ;\,m,\,\mu,q_4\right)\buildrel{ |x| \to
\infty}\over={ 1 \over ix } \sum_{r=0}^{\infty}b_r(m,\,\mu,q_4)
+ { 1 \over (ix)^2 } \sum_{r=0}^{\infty}\left[ a_r(m,\,\mu,q_4)
-  r \, b_r(m,\,\mu,q_4) \right] + {\cal O}\left({ 1 \over x^3
}\right)\, ,\cr \cr
&&f_1\left( x ;\,m,\,\mu,q_4\right)\buildrel{ |x| \to
\infty}\over={ 1 \over ix } \sum_{r=0}^{\infty} a_r(m,\,\mu,q_4)
- { 1 \over (ix)^2 } \sum_{r=0}^{\infty}
r \; a_r(m,\,\mu,q_4) + {\cal O}\left({ 1 \over x^3
}\right)\, .
\eea

For  $ Q^{(3)}\left( x ;\,m,\,\mu,q_4\right) $ we find the conditions,
\bea \label{conT}
&& \alpha_1(m,\mu,q_4) \; \sum_{r=0}^{\infty} b_r(m,\,\mu,q_4) +
\alpha_2(m,\mu,q_4) \; \sum_{r=0}^{\infty}  a_r(m,\,\mu,q_4) = -
\sum_{r=0}^{\infty} c_r(m,\,\mu,q_4) \, ,\cr \cr
&&\alpha_1(m,\mu,q_4) \; \sum_{r=0}^{\infty} \left[ a_r(m,\,\mu,q_4)
-  r \, b_r(m,\,\mu,q_4) \right] - \alpha_2(m,\mu,q_4) \;
\sum_{r=0}^{\infty} r \; a_r(m,\,\mu,q_4) =\cr \cr
&&= \sum_{r=0}^{\infty}\left[ r \,c_r(m,\,\mu,q_4)
-   b_r(m,\,\mu,q_4)\right] \; .
\eea
Imposing the same constraint on  $ Q^{(2)}\left(
x;\,m,\,\mu,q_4\right) $ yields,
\bea\label{conD}
&&\gamma_1(m,\mu,q_4) \; \sum_{r=0}^{\infty}  a_r(m,\,\mu,q_4) -
\gamma_2(m,\mu,q_4) \; \sum_{r=0}^{\infty}  a_r(m,-\mu,q_4)=
- \sum_{r=0}^{\infty} b_r(m,\,\mu,q_4)  \, ,\cr \cr
&&\gamma_1(m,\mu,q_4) \; \sum_{r=0}^{\infty}  r \, a_r(m,\,\mu,q_4) +
\gamma_2(m,\mu,q_4) \; \sum_{r=0}^{\infty} r \, a_r(m,-\mu,q_4) =\cr \cr
&& =
\sum_{r=0}^{\infty}\left[  a_r(m,\,\mu,q_4) - r \, b_r(m,\,\mu,q_4)\right]
\eea
Eqs.(\ref{conT}) can be easily solved
yielding,
\bea
&&\alpha_1(m,\mu,q_4) = { 1 \over
\Delta_{\alpha}(m,\mu,q_4)}\left\{\sum_{r=0}^{\infty} r \; a_r(m,\,\mu,q_4)
\sum_{n=0}^{\infty} c_n(m,\,\mu,q_4)  \right. \cr \cr
&&\left. -\sum_{n=0}^{\infty} a_n(m,\,\mu,q_4)
\sum_{r=0}^{\infty}\left[ r \,c_r(m,\,\mu,q_4)
-   b_r(m,\,\mu,q_4)\right]\right\} \, ,\cr \cr
&& \alpha_2(m,\mu,q_4) = { 1 \over \Delta_{\alpha}(m,\mu,q_4) }
\left\{ \sum_{n=0}^{\infty} a_n(m,\,\mu,q_4)
\sum_{r=0}^{\infty}\left[ a_r(m,\,\mu,q_4) - r \,
 b_r(m,\,\mu,q_4)\right] \right.  \cr \cr
&&\left. -\sum_{r=0}^{\infty} b_r(m,\,\mu,q_4) \sum_{n=0}^{\infty} n
\; a_n(m,\,\mu,q_4) \right\}
\eea
where
$$
\Delta_{\alpha}(m,\mu,q_4) \equiv \sum_{n=0}^{\infty} a_n(m,\,\mu,q_4)
\sum_{r=0}^{\infty} \left[ r \, b_r(m,\,\mu,q_4) - a_r(m,\,\mu,q_4)\right]
- \sum_{r=0}^{\infty} b_r(m,\,\mu,q_4) \sum_{n=0}^{\infty} n \;
 a_n(m,\,\mu,q_4) \; .
$$
We analogously obtain from  eqs. (\ref{conD})
\bea
&&\gamma_1(m,\mu,q_4) = { 1 \over \Delta_{\gamma}(m,\mu,q_4)}
\left\{\sum_{n=0}^{\infty} a_n(m,-\mu,q_4)\sum_{r=0}^{\infty}\left[
a_r(m,\mu,q_4) - r \,  b_r(m,\mu,q_4) \right]\right. \cr \cr
&&\left. - \sum_{n=0}^{\infty}
b_n(m,\mu,q_4)\sum_{r=0}^{\infty}  r \,a_r(m,-\mu,q_4)\right\} \; ,\cr \cr
&&\gamma_2(m,\mu,q_4) = { 1 \over \Delta_{\gamma}(m,\mu,q_4)}
\left\{\sum_{n=0}^{\infty} a_n(m,\mu,q_4)\sum_{r=0}^{\infty}\left[
a_r(m,\,\mu,q_4) - r \, b_r(m,\,\mu,q_4)\right]\right. \cr \cr
&&\left. +\sum_{n=0}^{\infty}
b_n(m,\mu,q_4)\sum_{r=0}^{\infty}  r \,a_r(m,\mu,q_4)\right\} \; ,
\eea
where,
$$
\Delta_{\gamma}(m,\mu,q_4) \equiv \sum_{n=0}^{\infty} a_n(m,\mu,q_4)
\sum_{r=0}^{\infty}  r \,a_r(m,-\mu,q_4)
+ \sum_{n=0}^{\infty} a_n(m,-\mu,q_4)\sum_{r=0}^{\infty}  r
  \,a_r(m,\mu,q_4) \; .
$$

\section{
Linear Relations among the solutions of the Baxter
equation for the quarteton case.}

We present here a proof of the linear recurrence relations
(\ref{ecrecu}) for the quarteton case. The  Baxter solutions
$Q^{(3)}\left( x ;\,m,\,\mu,q_4\right)$ and
$ Q^{(2)}\left( x ;\,m,\,\mu,q_4\right) $ are related by
\bea\label{linrel}
&&\left[ \alpha_1(\mu,\,m,q_4) - \gamma_1(\mu,\,m,q_4) + \pi
\cot{\pi x}
\right] Q^{(2)}\left( x ;\,m,\,\mu,q_4\right) \cr \cr
&&= Q^{(3)}\left(
x ;\,m,\,\mu,q_4\right) - \gamma_2(\mu,\,m, q_4) Q^{(2)} \left(
- x ;\,m,\,-\mu,q_4\right) \; .
\eea
This is eq.(\ref{ecrecu}) for $ n=4, \; r=2 $ and $
\delta^{(2)}(\mu,\,m,q_4) =
\alpha_1(\mu,\,m,q_4) - \gamma_1(\mu,\,m,q_4) , \; \alpha^{(2)}(\mu,\,m,q_4)
=
 - \gamma_2(\mu,\,m, q_4) $. Notice that $ Q^{(1)}\left( x
;\,m,\,\mu,q_4\right) $ is proportional to $ Q^{(2)}\left( -x
;\,m,\,-\mu,q_4\right) $ according to eq.(\ref{relsim}) and therefore
eq.(\ref{linrel}) is the only independent three-terms linear relation
between Baxter solutions.

In order to prove this relations we compute its triple, double and
simple poles at $  x = l \in {\cal Z} $. In the course of these
calculations we use the fact, that $ \pi \cot{\pi x} $
as a function of $ x $ has unit residues in all these  poles.

Moreover, we have from eqs.(\ref{elem4}) and (\ref{solu4}) that
\bea\label{AA}
&&Q^{(2)}\left( x ;\,m,\,\mu,q_4\right) \buildrel{ x \to
r\geq 0}\over=
{a_r(m,\,\mu,q_4) \over (x-r)^2}+{b _r(m,\,\mu,q_4) +
\gamma_1(\mu,;\,m,q_4) \; a_r(m,\,\mu,q_4) \over
x-r } \cr \cr
&&+ A_r(m,\,\mu,q_4) + {\cal O}\left(x-r \right)
\cr \cr
&&Q^{(2)}\left( x ;\,m,\,\mu,q_4\right) \buildrel{ x \to
- r<0}\over=
- \gamma_2(\mu,\,m, q_4) \; {a_r(m,\,-\mu,q_4) \over x+r}
+ {\hat A}_r(m,\,\mu,q_4)  + {\cal O}\left(x+r \right) \nonumber \\
\eea

Inserting eqs.(\ref{AA}) into the Baxter equation (\ref{eqBa4}) yields the
recurrence relations

\bea
&&
(r+1)^4 \,A_{r+1}(m,\,\mu,q_4)= \left[ 2\,r^4+m(m-1)\,r^2-\mu\, r+ q_4
\right]\,
A_r(m,\,\mu,q_4)-(r-1)^4\,A_{r-1}(m,\,\mu,q_4) \cr\cr
&&
+\left[ 8\,r^3+2m(m-1)\, r -\mu \right]\, b_r(m,\,\mu,q_4)-
4\,(r+1)^3\,b_{r+1}(m,\,\mu,q_4)-4\,(r-1)^3\,b_{r-1}(m,\,\mu,q_4) \cr\cr
&&
- 6(r+1)^2 \, a_{r+1}(m,\,\mu,q_4) - 6(r-1)^2 \, a_{r-1}(m,\,\mu,q_4) + [12
r^2
+ m(m-1) ]a _r(m,\,\mu,q_4)
\cr\cr
&&+ \gamma_1(\mu,;\,m,q_4) \left\{ \left[ 8\,r^3+2m(m-1)\, r -\mu \right]\,
a_r(m,\,\mu,q_4) \right. \cr\cr
&& \left. -
4\,(r+1)^3\,a_{r+1}(m,\,\mu,q_4)-4\,(r-1)^3\,a_{r-1}(m,\,\mu,q_4) \right\}
\; .
\eea
and
\bea
&&
(r+1)^4 \,{\hat A}_{r+1}(m,\,\mu,q_4)= \left[ 2\,r^4+m(m-1)\,r^2+\mu\, r+
q_4
\right]\,
{\hat A}_r(m,\,\mu,q_4)-(r-1)^4\,{\hat A}_{r-1}(m,\,\mu,q_4) \cr\cr
&&
+\gamma_2(\mu,;\,m,q_4) \left\{\left[ 8\,r^3+2m(m-1)\, r +\mu
\right]\, a_r(m,\,-\mu,q_4) \right. \cr \cr
&&\left. -
4\,(r+1)^3\,a_{r+1}(m,\,-\mu,q_4)-4\,(r-1)^3\,a_{r-1}(m,\,-\mu,q_4)
\right\}  \; .
\eea
These recurrence relations can be solved in terms of the coefficients
$a_r, \; b_r$ and $c_r$ using eq.(\ref{rr4}). One has to take into account
that $ A_r(m,\,\mu,q_4) $ and $ {\hat A}_r(m,\,\mu,q_4) $ do not obey
the initial conditions (\ref{condin}). We find,
\bea\label{solrec}
&& A_r(m,\,\mu,q_4) = c_r(m,\,\mu,q_4) + \gamma_1(\mu,;\,m,q_4) \;
 b_r(m,\,\mu,q_4) + A_0(m,\,\mu,q_4) \; a_r(m,\,\mu,q_4) \cr\cr
&&{\hat A}_r(m,\,\mu,q_4) = \gamma_2(\mu,;\,m,q_4) \;
b_r(m,\,-\mu,q_4) + A_0(m,\,\mu,q_4) \; a_r(m,\,-\mu,q_4)
\eea

Using eqs.(\ref{elem4}), (\ref{solu4}) and (\ref{solrec}) we find that
eq.(\ref{linrel}) holds at all of its poles. Therefore, the
r. h. s. and the l. h. s. of eq.(\ref{linrel}) can only differ on an
entire function. Taking into account the asymptotic behaviour of the
Baxter functions we conclude that this entire function is identically
zero.

\end{document}